\documentclass[twocolumn,epjc3]{svjour3}  

\usepackage[T1]{fontenc} 
\usepackage{epsfig}  
\usepackage{graphicx}  
\usepackage{amssymb}
\usepackage{amsmath}
\usepackage{pifont}
\usepackage[english]{babel}
\usepackage{color}  
\definecolor{darkgreen}{rgb}{0.2,0.5, 0.2}

\usepackage{dcolumn,booktabs}
\newcolumntype{d}[1]{D{.}{.}{#1}}
\usepackage{lipsum}
\usepackage{cite}
\usepackage{multirow}

\usepackage{dsfont}

\begin{document}

\title{Skyrme-Hartree-Fock-Bogoliubov mass models on a 3D mesh: 
       IIb. Fission properties of BSkG2.}

%% Group authors per affiliation:
%\author{Wouter Ryssens$^{1}$}
%\email{wryssens@ulb.be}
%\author{Guillaume Scamps$^{1}$}
%\author{Stephane Goriely$^{1}$} 
%\author{Michael Bender$^{2}$}
%\address{$^{1}$Institut d'Astronomie et d'Astrophysique, Universit\'e Libre de Bruxelles, Campus de la Plaine CP 226, 1050 Brussels, Belgium}
%\address{$^{2}$Universit{\'e} de Lyon, Universit{\'e} Claude Bernard Lyon 1, CNRS, IP2I Lyon / IN2P3, UMR 5822, F-69622, Villeurbanne, France}

\author{ Wouter Ryssens  \thanksref{e1,addr1}
        \and
        Guillaume Scamps \thanksref{addr1,addr2}
        \and
        Stephane Goriely \thanksref{addr1}
        \and
        Michael Bender   \thanksref{addr3}    
}
\thankstext{e1}{e-mail: wouter.ryssens@ulb.be}

\institute{Institut d'Astronomie et d'Astrophysique, Universit\'e Libre de Bruxelles, Campus de la Plaine CP 226, 1050 Brussels, Belgium \label{addr1}
           \and
           Department of Physics, University of Washington, Seattle, Washington 98195-1560, USA \label{addr2}
           \and
           Universit{\'e} de Lyon, Universit{\'e} Claude Bernard Lyon 1, CNRS / IN2P3, IP2I Lyon, UMR 5822, F-69622, Villeurbanne, France \label{addr3}
}

\maketitle

\begin{abstract} 
Large-scale models of nuclear structure are currently the only way to provide 
consistent datasets for the many properties of thousands of exotic nuclei 
that are required by nucleosynthesis simulations. 
In [W. Ryssens et al., Eur. Phys. J. A \textbf{58}, 246 (2022)], we recently 
presented the new BSkG2 model based on an energy density functional of the 
Skyrme type. Relying on a flexible three-dimensional coordinate representation
of the nucleus, the model takes into account both triaxial deformation and 
time-reversal symmetry breaking. BSkG2 achieves a state-of-the-art
global description of nuclear ground state (g.s.) properties and reproduces in particular
the known masses with a root-mean-square (rms) deviation of 678 keV. 
Moving beyond g.s. properties, the model also reproduces all empirical 
values for the primary and secondary barriers as well as isomer
excitation energies of actinide nuclei with rms deviations below 500 keV, 
i.e. with unprecedented accuracy. Here we discuss in detail the extension 
of our framework to the calculation of the fission barriers of 45 actinide nuclei, 
including odd-mass and odd-odd systems. We focus in particular on the impact of symmetry breaking which is
key to the accuracy of the model: we allow systematically
for axial, reflection and time-reversal symmetry breaking. The effect of the latter on the fission 
properties of odd-mass and odd-odd nuclei is small, but we find that allowing 
for shapes with triaxial or octupole deformation, as well as shapes with 
\emph{both}, is crucial to achieving this accuracy. The numerical accuracy of our
coordinate space approach, the variety of nuclear configurations explored and
the simultaneous successful description of fission properties and known masses
makes BSkG2 the tool of choice for the large-scale study of nuclear structure.
\end{abstract}

%%%%%%%%%%%%%%%%%%%%%%%%%%%%%%%%%%%%%%%%%%%%%%%%%%%%%%%%%%%%%%%%%%%%%%%%%%%%%%%%
\section{Introduction}

Nuclei play a prominent role in astrophysics: their reactions and decays release the 
energy that powers the light emitted by stars, staving off their eventual collapse
for as long as these processes can be maintained. 
The few hundred species of stable nuclei occurring on Earth
can teach us much, but not everything: thousands of short-lived isotopes play
a role in the nucleosynthesis of heavy elements and the structure of exotic 
astrophysical objects such as neutron stars. Understanding these aspects of the Universe
requires data on the properties of these nuclei, which can in turn be used
to model their reactions and decays. The most demanding application in terms
of data requirements is the simulation of the rapid neutron capture process or r-process, a 
process which produces a sequence of increasingly neutron-rich elements through 
repeated neutron captures which decay through a variety of channels to produce 
stable elements heavier than iron~\cite{Arnould20}. 

It is impossible to measure all relevant properties of the enormous number
of exotic nuclei involved in the r-process, the majority of which have so far 
not even been synthesised on Earth. The only way to access the required information then is modelling: nuclear theory
endeavours to construct large-scale models that can provide reliable 
extrapolations for neutron-rich nuclei. The focus of this effort is often nuclear 
binding energies: semi-empirical models, either based on the original work by 
Von Weizs{\"a}cker~\cite{Weizsacker35} or that by Duflo and Zuker~\cite{Duflo95}, 
microscopic-macroscopic (`mic-mac') approaches such as those of 
Refs.~\cite{Moller88,Moller95,Moller16} and microscopic approaches based on energy density functionals 
(EDFs)~\cite{Tondeur78,Goriely09b,Goriely16,Pena16,Scamps21,Ryssens22}. 
Most of these models reach a root-mean-square (rms) deviation for all nuclear 
masses that ranges from 500 to 800 keV, though some of the semi-empirical 
formulas can reach 200-300 keV~\cite{Duflo95}. Machine learning 
techniques have been used to produce standalone models of nuclear masses~\cite{Yuksel21}, 
but are more often used to augment the other approaches 
mentioned~\cite{Gazula92,Wang11,Wang14,Utama17,Neufcourt18,Shelley21,Ziu22}. 
Although such hybrid approaches can sometimes reach an extremely small rms deviation on the 
known masses of 100 keV or lower~\cite{Ziu22}, the resulting extrapolations towards neutron-rich 
nuclei still depend on the underlying model~\cite{Ye22}.

However, nucleosynthesis simulations require a more complete picture of nuclear 
structure than just an accurate description of masses; what is needed is 
a consistent set of all nuclear structure properties needed for reaction 
and decay calculations. Fission properties for instance, impact several aspects 
of the r-process such as (i) the details of fission recycling, (ii) the r-process abundances in the $110 \leq A \leq 170$ region, 
(iii) the production of cosmic chronometers such as 
Th and U~\cite{Fowler60,Goriely01} and (iv) the heating rate of kilonovae~\cite{Goriely15c}.
The theoretical description of nuclear fission in general and the fission of 
actinides in particular has a long history~\cite{Bjornholm80} and remains
today a very active field of research, see e.g. Refs.~\cite{Krappe12a,Schunck16a,Schmidt18a,Schunck22a} for 
recent reviews. Nevertheless, the community still faces many challenges~\cite{Bender20} 
and a complete description of fission remains difficult, particularly when 
aiming at thousands of neutron-rich nuclei. This difficulty partially explains the relative rarity of large-scale fission models when compared to the available mass models. 
Models that combine an accurate reproduction of masses and fission properties 
are even more rare: only microscopic-macroscopic approaches~\cite{Moller16} and EDF-based
models~\cite{Mamdouh01,Giuliani13,Goriely16} have been succesful at describing 
both simultaneously\footnote{For microscopic-macroscopic models an accurate 
description of fission barriers deteriorates the description of the masses, 
hence why Ref.~\cite{Moller16} advocates using two separate models: FRDM for 
masses and FRLDM for fission.}. Of the two, only EDFs offer a consistent 
microscopic description of the nucleus in terms of its constituent nucleons. 

Employing EDFs of the Skyrme type~\cite{Bender03}, we have recently 
started the development of the Brussels-Skyrme-on-a-Grid (BSkG) series of 
models~\cite{Scamps21,Ryssens22} in the spirit of the earlier BSk-models 
(see Ref.~\cite{Goriely16} and references therein). The main goal of the BSkG 
models is providing consistent nuclear data for astrophysical applications  
that is competitive with more phenomenological approaches, while gradually 
including additional physical ingredients in a microscopic way. Key to this 
strategy is the inclusion of thousands of nuclear
binding energies in the objective function of the parameter adjustment, which 
imposes both harsh constraints on the parameter values and requires significant
computational effort. Where the BSk-models relied on the assumption of axial
symmetry that reduces the structure of a nucleus to two dimensions, the BSkG-series
does away with this restriction and utilises a numerical representation in 
terms of a three-dimensional coordinate space mesh~\cite{Ryssens16}. 
In Ref.~\cite{Scamps21} we employed this representation to fit the parameters
of BSkG1 to the masses of thousands of nuclei, all of which were 
free to exploit triaxial deformation to lower their total energy.
For the construction of BSkG2, we also included degrees of freedom that break
time-reversal symmetry in the nuclear mean fields: these impact the masses of odd-mass and 
odd-odd nuclei and several other quantities such as magnetic moments and rotational
moments of inertia~\cite{Ryssens22}. 

Here we extend the reach of the BSkG series to nuclear fission: we study in 
detail the fission barriers and isomer excitation energies predicted by BSkG2.
Unfortunately, a good fit to ground state (g.s.) properties does not imply 
realistic predictions for barriers, as nuclear ground states exhibit small to 
moderate deformation while the shapes relevant to fission are
typically very elongated~\cite{Goriely07,Kortelainen15,Jodon16}. 
For this reason we included in the objective function of the BSkG2 model the 
RIPL-3 empirical values for the primary and secondary fission barriers of
twelve even-even actinide nuclei~\cite{Capote09} as well as the fission isomer 
excitation energies for seven of these. As already discussed briefly in 
Ref.~\cite{Ryssens22}, BSkG2 achieves an unprecedented accuracy for three 
different fission properties: primary and secondary barrier
heights and isomer excitation energies. For the 45 nuclei with $Z \geq 90$ 
 included in the RIPL-3 database~\cite{Capote09}, the model achieves 
rms deviations below 500 keV for all three quantities \emph{simultaneously}.
This accuracy does not come at the expense of 
masses: BSkG2 describes the 2457 known binding energies of  nuclei with $Z\geq 8$ 
in AME20~\cite{Wan21} with an rms of 678 keV.

Aside from the lowered rms deviations as compared to earlier models, 
there are two further points that make BSkG2 uniquely suited to the large-scale
description of fission. The first is our use of a coordinate-space representation, which 
allows us to reach high numerical accuracy even for the extremely elongated 
shapes associated with nuclear fission~\cite{Ryssens15b}.
The second is the large diversity of nuclear configurations we consider: we consistently include triaxial deformation in all our 
fission calculations and, where relevant, combine it with reflection asymmetry 
and octupole deformation. For the double-humped barriers of actinide nuclei, 
triaxial deformation can lower both the inner barrier by more than an MeV~\cite{Girod83,Bender03,Abusara10} 
and the outer barrier by several hundred keV~\cite{Lu2014,Ryssens19b,Ling20}.
Although we do not study any superheavy nuclei with $Z \gtrsim 110$ here, 
their fission barriers can be even more strongly affected~\cite{Cwiok96,Bender98,Warda02,Abusara12}. 
Despite this, triaxial deformation has up to now been neglected in large-scale 
studies of fission based on EDFs, likely due to its numerical cost. 
Furthermore, our treatment of the fission properties of odd-mass and odd-odd 
nuclei is as advanced as our description of their g.s. properties. 
Invoking no approximations, we describe such nuclei with blocking calculations
that include the effect of the time-odd terms of the EDF, although we will 
establish below that their  effect of the latter on 
barriers and isomer excitation energies is small. To the best of our knowledge,
we are the first to report on self-consistent Hartree-Fock-Bogoliubov (HFB) 
calculations that consider blocking for odd-mass and odd-odd nuclei while
breaking axial, reflection and time-reversal symmetry simultaneously. 

This paper is organized as follows: we summarize the relevant aspects 
of the construction of BSkG2 and specify the numerical conditions of the fission
calculations in Sec.~\ref{sec:model}. In Sec.~\ref{sec:Pu240} we discuss the 
fission barriers of $^{232}$U, $^{240}$Pu and $^{244}$Pu as representative
examples. We discuss fission properties of nuclei in the actinide region more systematically
in Sec.~\ref{sec:systematics} and present our
conclusions and outlook in Sec.~\ref{sec:conclusions}.

%%%%%%%%%%%%%%%%%%%%%%%%%%%%%%%%%%%%%%%%%%%%%%%%%%%%%%%%%%%%%%%%%%%%%%%%%%%%%%%%
\section{The BSkG2 model}
\label{sec:model}

%%%%%%%%%%%%%%%%%%%%%%%%%%%%%%%%%%%%%%%%%%%%%%%%%%%%%%%%%%%%%%%%%%%%%%%%%%%%%%%%
\subsection{Binding energy and collective correction}

We define the total binding energy of a nucleus, represented by a
state $| \Phi \rangle$ of the HFB type as:
\begin{align}
E_{\rm tot} &= E_{\rm HFB} + E_{\rm corr} \, .
\label{eq:Etot}
\end{align}
In Eq.~\eqref{eq:Etot}, $E_{\rm HFB}$ is the self-consistent mean-field 
energy and $E_{\rm corr}$ is the collective correction energy. The mean-field energy
contains the kinetic energy as well as the Skyrme and Coulomb energies that 
determine, respectively, the strong and electrostatic interaction between the 
nucleons. The form of the Skyrme EDF we employ is essentially standard, but
does include well-defined time-odd terms~\cite{Ryssens22}.

The goal of the correction energy is modelling correlations that cannot be
captured by a single HFB state built from separate proton and neutron 
orbitals: rotational and vibrational collective motion, centre-of-mass 
motion and proton-neutron pairing. These four effects each give rise to a term 
in the correction energy: 
\begin{align}
E_{\rm corr}&= E_{\rm rot} + E_{\rm vib} + E^{(2)}_{\rm cm} + E_{\rm W}  \, ,
\label{eq:Ecorr}
\end{align}
where these are, respectively, the rotational and 
vibrational correction~\cite{Tondeur00,Goriely13b}, the two-body part of the 
centre-of-mass correction~\cite{Bender00}, and the Wigner energy~\cite{Goriely03}. 
The rotational and vibrational correction have the same form and are combined as
%
%\begin{subequations}
\begin{align}
\label{eq:Erot}
E_{\rm rot} + E_{\rm vib} &= - 
\sum_{\mu=x,y,z} \left( f_{\mu}^{\rm rot} + f_{\mu}^{\rm vib} \right) \frac{\langle \hat{J}^{2}_{\mu} \rangle}{2 \mathcal{I}_{\mu}} \, ,
\end{align}
%\end{subequations}
%
where $\mathcal{I}_{\mu}$ is the Belyaev moment of inertia (MOI) around 
Cartesian axis $\mu$. The factors $f_{\mu}^{\rm rot/vib}$ are both defined in 
terms of $B_{\mu} = \mathcal{I}_{\mu} / \mathcal{I}_c$, the ratio of the 
Belyaev MOI to that of (one third of) a rigid rotor $\mathcal{I}_c$~\cite{Ryssens22}: 
\begin{subequations}
\begin{align}
f_{\mu}^{\rm rot} &=  b  \tanh \left( c B_{\mu} \right) \, , \\
f_{\mu}^{\rm vib} &= d B_{\mu} e^{- l \left( B_{\mu} - B_0 \right)^2} \, .
\end{align}
\end{subequations}
As in Ref.~\cite{Ryssens22}, we stress that the vibrational component of 
Eq.~\eqref{eq:Erot} only intends to capture the \emph{deformation dependence} 
of the spurious collective vibrational energy. Since Eq.~\eqref{eq:Erot}
vanishes for spherical configurations, we put the burden of simulating any
collective vibrational energy that is deformation independent on 
the EDF coupling constants.
Expressions for all other ingredients of the total energy can be found in 
Refs.~\cite{Scamps21,Ryssens22}.

%%%%%%%%%%%%%%%%%%%%%%%%%%%%%%%%%%%%%%%%%%%%%%%%%%%%%%%%%%%%%%%%%%%%%%%%%%%%%%%%
\subsection{Shapes, fission paths and barriers}
\label{sec:fission}

Modelling nuclear fission starts with the selection of a set of collective 
coordinates that characterize the large-scale collective motion of the nucleus 
on its way to scission. Traditionally, one uses a small number of (mass) 
multipole moments $Q_{\ell m}$ that characterize the shape of the nucleus. We 
define the $Q_{\ell m}$ and their dimensionless equivalents $\beta_{\ell m}$ 
for integer ($\ell , m$) with $\ell \geq 1$ and $0 \leq m \leq \ell$ as 
\begin{subequations}
\begin{align}
\langle Q_{\ell m} \rangle &=    
                  \int d^3r  \, \rho_0(\bold{r}) r^{\ell}  \Re \left[ Y_{\ell m}(\theta, \phi)\right] \, ,
\label{eq:qlm}\\
\beta_{\ell m } &= \frac{4 \pi } {  3  R^{\ell}  A } Q_{\ell m} \, ,
\label{eq:betalm}
\end{align}
\end{subequations}
where $R = 1.2 A^{1/3}$ fm. 
Replacing the total density $\rho_{0} (\boldsymbol{r})$ in Eq.~\eqref{eq:qlm} by the proton, neutron or
charge density and substituting $A$ by the appropriate particle number in the 
denominator of Eq.~\eqref{eq:betalm}, one obtains the deformations
$\beta_{\ell m, p}, \beta_{\ell m, n}$ or $\beta_{\ell m, c}$, respectively.
For plotting purposes, we will also employ the alternate $(\beta, \gamma)$ 
characterization of quadrupole deformation, defined as
\begin{subequations}
\begin{align}
\label{eq:beta}
\beta  &= \sqrt{\beta_{20}^2 + 2 \beta_{22}^2 } \, , \\
\gamma &= \text{atan} \left( \sqrt{2}\beta_{22}/ \beta_{20} \right) \, . 
\label{eq:gamma}
\end{align}
\end{subequations}
One of the quadrupole deformations, $\beta_{20}$, characterizes the elongation 
of the nuclear shape in the $z$-direction and is virtually always employed as 
collective coordinate. The octupole deformation, $\beta_{30}$, reflects the 
left-right asymmetry of the nuclear density distribution along the $z$-axis and 
is a traditional second choice. The second quadrupole degree of freedom 
$\beta_{22}$~\cite{Abusara10,Sadhukhan14,Lu2014,Ling20}, 
the hexadecapole moment $\beta_{40}$~\cite{Warda12} and the particle number 
dispersions $\Delta N_q^2 \equiv \langle \hat{N}_q^2 \rangle - \langle \hat{N}_q \rangle^2$ 
of both nucleon species ($q=p,n$)~\cite{Sadhukhan14} have all been used as 
collective coordinates in the study of fission paths.

% Here there should be tunnelling
Once a set of coordinates has been chosen, the goal is to study the movement of 
the nucleus in this collective space along (continuous) paths that connect 
the ground state to a configuration of two disconnected nuclei, i.e.\ a fissioned
system. By minimizing an appropriate action integral, one can obtain the 
trajectory that gives the nucleus the highest probability of tunneling through 
the barrier(s) in a semiclassical approach~\cite{Baran81,Lemaitre18}: 
this is the least action path (LAP). The action integral depends on the total energy and the 
inertial mass tensor associated with the collective coordinates, and the calculation
of both in the entire collective space is required in order to be able to minimize the action. 

Ideally, one (i) constructs a complete potential energy surface (PES) in all 
relevant collective degrees of freedom and (ii) minimizes the action among all 
possible paths. Both of these represent enormous computational challenges 
if one moves beyond more than a few collective degrees of freedom. 
This study is an initial step towards more extensive calculations for thousands 
of nuclei; in view of this computational challenge, we limit ourselves here to two 
collective degrees of freedom: $\beta_{20}$ and $\beta_{22}$.
We do not attempt to construct the LAP, but restrict ourselves to 
the simpler concept of the least energy path (LEP)~\cite{Lemaitre18}. This 
path can be constructed without calculation of the inertia tensor,
and therefore ignores all dynamical aspects of the fission process.
The local maxima encountered along the (one-dimensional) LEP are, by 
construction, saddle points on the complete (multi-dimensional) PES. It is 
the energies of these saddle points, normalized to the corresponding g.s., that 
we compare to the empirical values of the RIPL-3 database.

Our calculation of the fission properties of a nucleus thus consists of two steps.
First, we explore the relevant part of the collective space with numerous 
EDF calculations constrained to different values of the collective 
coordinates $(\beta_{20}, \beta_{22})$. A robust algorithm to adjust the 
numerical parameters of the quadratic constraints we imposed on these multipole 
moments greatly simplifies these calculations~\cite{Ryssens16}. The shape 
degrees of freedom explored by nuclei in our calculations are however \emph{not}
restricted to the two quadrupole moments: a self-consistent EDF calculation will
employ all multipole moments to optimize the mean-field energy, limited only
by symmetries imposed on the calculation by either the numerical implementation 
or the starting point of the iterative process.
In constrast to microscoscopic-macroscopic approaches where the shape degrees
of freedom explored by the nuclei and the collective coordinates are identical, 
our calculations thus naturally include all multipole moments that our numerical
choices allow for, see also Sec.~\ref{sec:numerical} and Ref.~\cite{Ryssens22}.
More specifically, our calculations allow for non-zero values of the 
$\beta_{\ell m}$ for arbitrary $\ell > 0$ and even values of $m$\footnote{As for the g.s. calculations of Ref.~\cite{Ryssens22}, we still
impose z-signature as self-consistent symmetry which implies that $\beta_{\ell m}$
vanishes if $m$ is odd.}.
In particular, this set includes multipole moments with odd $\ell$ that break 
reflection symmetry, the most important of which is the octupole moment $\beta_{30}$.
Octupole deformation is crucial to the description of the fission of actinide
nuclei, but the construction of three-dimensional PESes in ($\beta_{20}, \beta_{22}, \beta_{30}$) 
is still prohibitively expensive for more than a handful of nuclei. 
Luckily, the topographies of the PESes of the nuclei we consider here 
are all similar and not too complicated: we will demonstrate in Sec.~\ref{sec:Pu240} 
that the LEPs we construct from two-dimensional PESes are closely equivalent to 
those we would have obtained from much more computationally expensive 
three-dimensional PESes.

Not all values of multipole moments correspond to distinct shapes though:
the (body-fixed) principal axes of a deformed but reflection-symmetric shape 
can be assigned to the $x$-, $y$- and $z$-axes in the simulation volume in three
different ways. If the configuration is invariant under time-reversal, 
these three possibilities are all physically equivalent. 
If the shape is in addition also reflection asymmetric, the number
of such equivalent orientations rises to six. When the configuration 
breaks time-reversal symmetry, all these possibilities are not exactly equivalent
any longer since the finite angular momentum determines a preferred direction in
space; the energy difference due to a reorientation of the principal axes 
has never been studied for odd-odd systems but is on the order of 100 keV
for heavy odd-mass nuclei~\cite{Schunck10}.
For simplicity, we ignore this reorientation effect entirely as we did
in Ref.~\cite{Ryssens22} and limit our calculations to a specific part of the 
collective space with little loss of 
generality: $\gamma \in [0, 60]^{\circ}$ and 
$\beta_{30} \geq 0$. Finally, the mass dipole ($\ell = 1$) moments are 
proportional to the center-of-mass of the nucleus and do not represent
physical degrees of freedom. Only $\beta_{10}$ is not restricted by 
our symmetry choices; we constrain it to be zero in all calculations in order
to optimize the placement of the nucleus in the simulation volume.

As second step to calculate the fission properties of a nucleus we use a
flooding model~\cite{Lemaitre18} to determine the LEP; along which we 
interpolate to obtain the saddle points. We determine fission isomer excitation 
energies by searching for the local minimum of the PES at moderate deformation, 
$\beta_{20} \sim 0.9$. 
This workflow is universally applied to all nuclei. Our treatment of  odd-mass 
and odd-odd nuclei differs from that of even-even nuclei only in the underlying 
EDF calculations: the PES was constructed from self-consistent calculations 
with quasiparticle blocking that included the effects of time-reversal symmetry breaking~\cite{Ryssens22}. 
We stick to the techniques of Ref.~\cite{Ryssens22}: we limit ourselves to 
quasiparticle excitations with $z$-signature eigenvalue $\eta = +i$ and
use a gradient-based HFB solver to combat convergence issues and guarantee that 
we obtain the lowest possible energy for each value of $(\beta_{20}, \beta_{22})$.

Fission paths obtained from self-consistent calculations with a limited number
of collective degrees of freedom can be plagued with discontinuities~\cite{Dubray12}. 
In this sense, our search for the LEP is crude: we made no attempt 
to explicitly check that the paths we obtain are continuous. Visual 
inspection of our results did not reveal obvious signs of discontinuities, 
such as large jumps in the total energy or any multipole moment from one point 
on the fission path to the next. It is not excluded that the LEPs we obtain are 
affected by more subtle discontinuities, but it is not 
easy to identify such problems and even harder to cure them if found.
We interpret the apparent absence of discontinuities as due to the inclusion
of triaxial deformation.
Most calculations of fission barriers that can be found in the 
literature restrict the nucleus to axial symmetry which separates all 
single-particle states into symmetry blocks determined by the $K$ quantum number. 
Triaxial deformation allows the nucleus 
to connect these symmetry blocks in a continuous way, drastically reducing the 
possibility for discontinuous changes in single-particle configuration from 
one calculated point on the PES to the next as compared to axially symmetric 
calculations. 

The situation is more complicated for odd-mass and odd-odd nuclei: the properties of the blocked quasiparticle(s) 
can vary dramatically for even small changes in deformation, introducing 
in this way a possible second source of discontinuities.
As in previous studies~\cite{Goriely07}, we simply take the overall
lowest energy at any grid point on the PES and thus pass over any consideration of conserved quantum numbers when 
establishing the LEP. Anticipating that we will find that all nuclei studied 
here will take triaxial shapes at all relevant saddle points for the LEP, our
choice can be motivated by the inevitable mixing of 
quasiparticles with different $K$ quantum numbers when the fission path passes 
through triaxial shapes, and the mixing of quasiparticles with different parity 
when passing through octupole-deformed shapes. In particular around the outer 
saddle point, $z$-signature is the only remaining quantum number of the 
single-particle states. All quasiparticles we consider for blocking purposes 
have $z$-signature $\eta = +i$, such that in this region of the path our 
calculations cannot encounter any level crossings that could introduce discontinuities.

Because the treatment of blocked quasiparticles adds significant complexity to
a calculation, fission barriers for odd-mass and odd-odd nuclei are rarely 
considered in the literature, with notable exceptions being 
Refs.~\cite{Perez09a,Heenen16,Koh17,Rodriguez17} and \cite{Schunck22x}. 
All of these studies are limited to a few nuclei only, and all limit the possible 
shapes to axial ones. The authors of these references also do not construct 
configurations with the lowest energy at each point of the collective space, 
but rather construct paths at fixed $K$ quantum number in order to reduce the occurrence of discontinuities encountered.
The large-scale fission calculations with BSk14 of Ref.~\cite{Goriely07} 
form an exception: while restricted to axial symmetry, the authors did not 
enforce constant $K$ along the fission path, likely leading to large numbers
of discontinuities. The question how to construct a consistent fission path in 
completely symmetry-unrestricted calculations will require further investigation
in the future~\cite{Bender20}, particularly when attempting the description
of more complex observables such as lifetimes or fission yields.

%%%%%%%%%%%%%%%%%%%%%%%%%%%%%%%%%%%%%%%%%%%%%%%%%%%%%%%%%%%%%%%%%%%%%%%%%%%%%%%%
\subsection{Parameter adjustment and global performance}
\label{sect:parameter:adjustment}

The total binding energy $E_{\rm tot}$ depends on 25 parameters that were
adjusted to experimental data. We reported the BSkG2 parameter values in 
Ref.~\cite{Ryssens22} and discussed there the parameter adjustment and the 
performance of the model for g.s. properties. Here, we summarize only
a few key points.

The main ingredient of the objective function is the ensemble of 2457 
measured binding energies ($Z\geq 8$) tabulated in the AME20 database~\cite{Wan21}. 
We also constrained the EDF parameters so that the calculated 
$uv$-averaged neutron pairing gaps $\langle \Delta \rangle_n$~\cite{Bender00b} 
reproduce as best as possible the experimental five-point
gaps $\Delta^{(5)}_n$ in order to obtain a realistic pairing strength for the neutrons. 
The fit included constraints on the models infinite nuclear matter properties, 
such as the symmetry energy ($J \in [30,32]$ MeV) 
the nuclear incompressibility ($K_{\nu} \in [230,250]$  MeV and the isoscalar 
effective mass $M_s^* / M \approx 0.84$). Finally, we adjusted the Fermi wave 
number $k_F$ to reproduce 884 measured charge radii~\cite{Angeli13}.

The selection of g.s. properties is close to those that figured in
the objective function of BSkG1. For BSkG2, we also included empirical values 
for the primary ($E_{\rm I}$) and secondary ($E_{\rm II}$)
fission barriers of twelve even-even nuclei with $92 \leq Z \leq 96$
from RIPL-3~\cite{Capote09} and their fission isomer excitation 
energies ($E_{\rm iso}$) when available~\cite{Samyn04}. As already explained in Ref.~\cite{Ryssens22}, 
we did not employ all nuclei included in the RIPL-3 database in order to 
(i) focus on nuclei with low fission barriers ($E_{\rm I} < 10$ MeV) due to their 
relevance to r-process nucleosynthesis and (ii) not to further complicate the 
adjustment process with the fission barriers of odd-mass and odd-odd nuclei.
Our final selection of twelve primary and secondary barrier heights and seven
isomer excitation energies is summarised in Table~\ref{tab:fission_fit_data}.
These empirical values are certainly not without associated uncertainties. 
The fission barriers themselves are at best pseudo-observables, quantities that 
can only be extracted in a model dependent
way. We are not aware of any attempt to quantify the uncertainties of the 
RIPL-3 recommended values. Although isomer excitation energies are directly
observable, their accurate measurement remains difficult: for example, the 
literature reports values for the isomer excitation energy of $^{240}$Pu that 
range from $2.25 \pm 0.2$ MeV~\cite{Hunyadi01} to $\approx 2.8$ MeV~\cite{Singh02}. 
This uncertainty also extends to the quantum numbers of the fission isomers: 
their spin and parity are not guaranteed to be identical to those of the 
ground state. Except for the fission isomer of $^{238}$U~\cite{Kantele83}, only a 
handful of tentative spin-parity assignments are available~\cite{Singh02}.

Two remarks on nomenclature are in order. First, we will follow much 
of the literature in using the word `barrier' as a shorthand for `excitation energy of the saddle point' 
to describe our results if the context permits us to do so. Second, for the 
double-humped fission barrier of actinide nuclei it is natural to 
discuss `inner' and `outer' barriers that correspond to the saddle points
encountered along the LEP at moderate and large elongation, 
respectively. `Primary barrier' and `secondary barrier'  then refer to the 
saddle points located at highest and lowest excitation energy compared to the g.s. minimum, 
respectively. The RIPL-3 database lists reference values for inner and 
outer barriers as deduced from fits to experimental fission cross sections. 
However, the fission transmission coefficients employed in such fits are not 
sensitive to the ordering of saddle points, only to their excitation energies~\cite{Capote09}. 
For this reason, our comparisons with empirical values in the objective 
function of the parameter adjustment and in the text below always concern 
primary and secondary barriers, \emph{not} inner and outer barriers.

%-------------------------------------------------------------------------------
% Vertical version of table
\begin{table}
\centering
\begin{tabular}{ccccc@{\hskip 10pt}cccc@{\hskip 10pt}cccc}
\hline
\hline
%\multicolumn{4}{c}{$^{A}_{92}$U}         \\ %& \multicolumn{4}{c}{$^{A}_{94}$Pu} & \multicolumn{4}{c}{$^{A}_{96}$Cm}  \\
 Z & A & $E_{\rm I}$ & $E_{\rm II}$ & $E_{\rm iso}$ \\ 
\hline
92 & 232 & 5.40 & 4.90 &  -       \\
92 & 234 & 5.50 & 4.80 &  -       \\
92 & 236 & 5.67 & 5.00 &  2.30    \\
92 & 238 & 6.30 & 5.50 &  2.60    \\ 
 \noalign{\smallskip}
94 & 238 & 5.60 & 5.10 &  2.40 \\
94 & 240 & 6.05 & 5.45 &  2.25 \\
94 & 242 & 5.85 & 5.05 &   -  \\
94 & 244 & 5.70 & 4.85 & 2.00  \\
 \noalign{\smallskip}
96 & 242 & 6.65 & 5.00 & 1.80\\
96 & 244 & 6.18 & 5.10 & 1.04 \\
96 & 246 & 6.00 & 4.80 & - \\
96 & 248 & 5.80 & 4.80 & - \\
\hline
\hline
\end{tabular}
\caption{ Values for the primary ($E_{\rm I}$) and secondary ($E_{\rm II}$) fission barriers 
          as well as the isomer excitation energies $E_{\rm iso}$ included in the parameter adjustment. 
          Barrier values are the empirical values recommended in Ref.~\cite{Capote09}, 
          isomer excitation energies are those of Ref.~\cite{Samyn04}.
          All energies are expressed in MeV.
          } 
\label{tab:fission_fit_data}
\end{table}
%-------------------------------------------------------------------------------

As explained in Sec.~\ref{sec:fission}, obtaining the fission path
of even a single nucleus is a computationally demanding and complex multi-step 
procedure that is difficult to automate consistently. This means that the practical inclusion 
of barriers in the objective function is extremely challenging and, 
to the best of our knowledge, has never been achieved for an EDF-based model before.
Instead, simplifications have been adopted: the adjustment of the 
early SkM$^*$ Skyrme parameterization employed semiclassical estimates of the 
barrier of $^{240}$Pu~\cite{Bartel82}, the surface coefficient of the Gogny D1S 
parameterization was manually changed to reproduce the barrier of 
$^{240}$Pu~\cite{Berger89} and the UNEDF1 and UNEDF2 parameterizations used
four fission isomer excitation energies instead of barriers in the adjustment
protocol~\cite{Kortelainen12,Kortelainen14}. The authors of Ref.~\cite{Jodon16} 
constructed a series of parameter sets with systematically varied values of the 
surface energy coefficient, of which SLy5s1 is the one that best describes the 
excitation energy of superdeformed states and fission barriers~\cite{Jodon16,Ryssens19b}.
The BSk14 parameterization was created in two steps: an initial fit to g.s. 
properties, followed by the fine-tuning of the (few) parameters characterizing 
the collective correction to fission barriers~\cite{Goriely07}.

For BSkG2, we adopted a two-step adjustment procedure identical to that used for
BSk14~\cite{Goriely07}. First, we adjusted all parameters of the EDF to the 
g.s. properties included in the objective function. Using the intermediate
parameter values issuing from this step, we calculated the PESes for the 
twelve even-even nuclei in Tab.~\ref{tab:fission_fit_data}.
Freezing all other parameters, we then adjusted the nine parameters of the 
collective correction ($V_W, \lambda, V'_W, A_0, b,c,d,l$ and $\beta_{\rm vib}$)
to the complete objective function including fission properties, 
g.s. properties and masses. Since the collective correction is treated 
semi-variationally, the changes induced in the PESes (and therefore the saddle 
points) by the variation of these parameters can be obtained at essentially no 
computational cost with the values of $\langle \hat{J}_{x/y/z}^2\rangle$ 
and $\mathcal{I}_{x/y/z}$ tabulated in the first step. We thus eliminated the 
need for the repeated construction of complete PESes, although we continued EDF 
calculations for the g.s.~properties of nuclei during the second step.

As already pointed out in \cite{Ryssens22}, this simple procedure for adjusting 
the barrier heights through a fine-tuning of the correction energy is sufficient 
because the fit to masses already provides quite realistic surface properties 
in the first step, with barriers typically deviating by less than 2 MeV from the 
empirical values. If the parameters issued from the first step yielded fission 
barriers that are too high by 10~MeV as is the case for many Skyrme 
EDFs~\cite{Jodon16}, this could not be corrected for by a small modification of a
correction whose variation between ground state and saddle points is typically 
on the order of 2 MeV, as we will illustrate in Sec.~\ref{sec:Pu240}. One of the 
keys to finding reasonable agreement for fission barriers without including 
them in the objective function is the inclusion of the two-body part of the 
centre-of-mass correction in the EDF~\cite{Bender00,daCosta22}, as we do here.

The resulting BSkG2 model achieves an rms deviation on the known nuclear masses of
$\sigma(M)$ = 0.678 MeV, and an rms deviation on the nuclear charge radii of 
$\sigma(R_c)$ =  0.0274 fm, while offering reasonable predictions for
the empirical properties of infinite nuclear matter~\cite{Ryssens22}. 
The performance of the model with respect to the AME20 masses is slightly better
than that of BSkG1, which has $\sigma(M)$ = 0.734 MeV~\cite{Scamps21}.  It is 
however worse than the later entries in the BSk series that have rms 
deviations somewhat below 0.6 MeV~\cite{Goriely16}. Nevertheless, BSkG2 does 
significantly better than other Skyrme models: they often reach rms deviations 
of more than two MeV~\cite{Klupfel09,Kortelainen10} as their fit protocol 
typically includes only the binding energies of a handful of nuclei.

We reported already in Ref.~\cite{Ryssens22} on the performance of the model 
on the fission properties of all 45 nuclei with $90 \leq Z$ in the RIPL-3 
data base~\cite{Capote09}, which reach up to $Z=96$.
BSkG2 reproduces both the primary and secondary barrier heights with high accuracy:  
the rms deviations are 0.44 MeV and 0.47 MeV, respectively. This accuracy 
is comparable to that achieved for 28 isomer excitation energies, which are described
with an rms deviation of $0.49$ MeV. Anticipating a more detailed comparison  
with other models in Sec.~\ref{sec:comparison}, we already mention here 
that this degree of accuracy for all three fission quantities is unrivalled in 
the existing literature.

%%%%%%%%%%%%%%%%%%%%%%%%%%%%%%%%%%%%%%%%%%%%%%%%%%%%%%%%%%%%%%%%%%%%%%%%%%%%%%%%
\subsection{Numerical representation and symmetries}
\label{sec:numerical}

As in Refs.~\cite{Scamps21,Ryssens22}, we use the MOCCa code for all EDF
calculations~\cite{Ryssens16}. It iterates $N_N$ neutron and $N_Z$ proton 
single-particle wave functions on a cubic three-dimensional Cartesian Lagrange 
mesh characterized by three numbers of discretisation points $N_x, N_y, N_z$ 
and a grid spacing $dx$~\cite{Baye86}. This coordinate representation is 
particularly well suited to the description of fission: it offers a numerical 
accuracy independent of the nuclear shape\cite{Ryssens15b}. The extremely 
elongated shapes relevant to fission can be accurately represented, with the
only proviso that the simulation volume be sufficiently large.

We impose $z$-signature and $y$-time-simplex symmetries on the nuclear configuration 
for all calculations reported on here, allowing us to reduce the effective
number of mesh points in $x$ and $y$-directions in the calculations by half
\cite{Ryssens16}. In regions of the PES where this lowered the total energy, 
we allow for octupole deformation by breaking reflection symmetry. This 
requires the explicit numerical representation of all mesh points in the 
$z$-direction but also an additional contraint to ensure the $z$-coordinate of the 
nuclear centre-of-mass, or equivalently $\beta_{10}$, vanishes at convergence.

For even-even nuclei, we imposed time-reversal symmetry such that we could 
restrict practical calculations to $(N_N + N_Z)/2$
single-particle wave functions. For odd-mass and odd-odd nuclei, we account 
for the effects of time-reversal symmetry breaking and explicitly 
represent all $N_N+N_Z$ single-particle states. As for the calculation of
g.s. properties~\cite{Ryssens22}, we limit ourselves to quasiparticle excitations
with $z$-signature $\eta = +i$ as mentioned in the previous section.

The most demanding calculations we report on here are those for odd-mass and 
odd-odd nuclei in regions of the PES where octupole deformation is relevant: 
elongated shapes require extended meshes and neither reflection nor time-reversal 
symmetry could be employed to simplify the calculations. To efficiently fit 
such calculations in the memory of the CPUs available to us, we employed a mesh 
with $N_z = 40$ and $N_x = N_y = 32$ points at $dx=0.8$ fm. This mesh is 
somewhat extended in the $z$-direction and reduced in the $x$ and $y$ directions
as  compared to our ground state calculations ($N_x = N_y = N_z = 36$). We used 
a limited number of single-particle states in the calculations: $N_N= 440$ neutron 
states and $N_Z = 260$ proton states, independently of the nucleus considered.
While these choices somewhat limit the numerical accuracy of our calculation of 
the \emph{absolute} energy as a function of deformation, we checked that the 
numerical error of any energy \emph{difference} is generally comparable to the 
error due to the mesh spacing, i.~e.\ about 100 keV~\cite{Ryssens15b}.
Furthermore, the outer barrier for actinide nuclei is typically located at 
$\beta_{20} \sim 1.3$, i.~e.\ at large, but not extreme, elongation of the nucleus.

%%%%%%%%%%%%%%%%%%%%%%%%%%%%%%%%%%%%%%%%%%%%%%%%%%%%%%%%%%%%%%%%%%%%%%%%%%%%%%%%
\section{Topography of an actinide fission barrier}
\label{sec:Pu240}

\begin{figure*}[t]
\centering
\includegraphics[width=.90\textwidth]{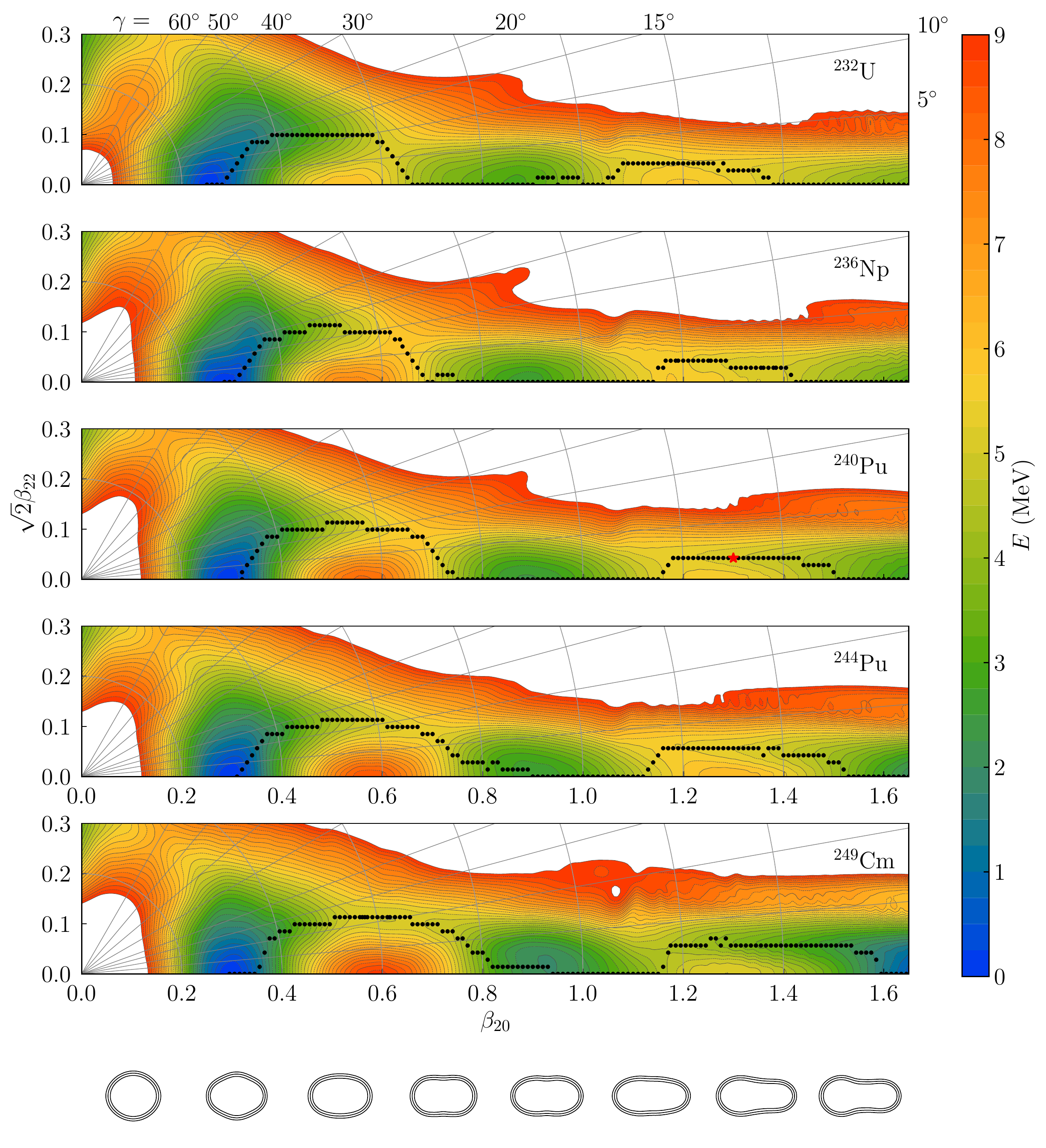}
\caption{ (Color online) 
          Potential energy surfaces as a function of quadrupole 
          deformation $\beta_{20}$ and $\sqrt{2}\beta_{22}$ for five
          nuclei. From top to bottom: $^{232}$U, $^{236}$Np, $^{240}$Pu, 
          $^{244}$Pu and $^{249}$Cm. All energies are normalized to the 
          g.s. minima near $\beta_{20} = 0.3$.
          Circles of constant total deformation $\beta$ and lines of constant 
          $\gamma$ are drawn in grey. Black circles indicate the lowest energy 
          fission path (see text). Contour lines in the five top 
          panels are 0.25 MeV apart. 
          An indication of the elongation and reflection asymmetry along the
          surface is given by the contour plots of the total density 
          of axially symmetric configurations of $^{240}$Pu 
          for $\beta_{20} = 0.1, 0.3, \ldots, 1.5$ (and $\beta_{22} = 0$)
          below the five panels. The red star in the middle panel indicates
          the location of the configuration drawn in Fig.~\ref{fig:fancy_3D}. 
          }
\label{fig:landscapes}
\end{figure*}

The PESes of the forty-five actinide nuclei we consider have comparable
topographies, which we will show explicitly in Sec.~\ref{sec:systematics}. 
Before discussing their systematics, we start here by discussing the typical 
features of PESes in this region of the nuclear chart using the example of
five isotopes that span nearly the entire range of $Z$ and $N$ of the 
empirical barriers:  $^{232}$U, $^{236}$Np, $^{244}$Pu, $^{249}$Cm and $^{240}$Pu. 
We will discuss the latter in more detail as it is regularly used as benchmark for 
fission studies \cite{Flocard74,Goriely07,Jodon16,Bartel82,Abusara10,Berger89,Rutz95,Bender03,Burvenich04,Bonneau04,Samyn05,Younes09,Li10,Abusara12,Schunck14,Ryssens19}.
The complete PESes of all five isotopes as a function of $\beta_{20}$ and 
$\sqrt{2}\beta_{22}$ are shown in Fig.~\ref{fig:landscapes}, with the LEP for each 
indicated by black circles. Axially symmetric (AS) prolate configurations 
correspond to points on the horizontal axes, while AS oblate configurations 
lie on the $\gamma=60^{\circ}$ lines. The five PESes show similar features 
such as a well-deformed AS prolate ground state near $\beta_{20} \sim 0.3$ and 
a second AS minimum near $\beta_{20} \sim 0.9$ that is more than 2.5 MeV above 
the less deformed minimum. Beyond the range of $\beta_{20}$ covered by the 
figure, the energy slopes smoothly down until the scission point is reached,
i.e.\ the deformation at which the nucleus breaks apart into two fragments. 
Three local maxima are visible in each panel: aside from the spherical 
point, one (inner) peak at $\beta_{20} \sim 0.55$ and another (outer) near 
$\beta_{20} \sim 1.3$ separate the g.s. and the isomeric state from scission 
configurations. For deformations larger than those of the isomer, actinide 
nuclei prefer reflection asymmetric shapes, characterized by an octupole 
moment which grows with increasing elongation. Since this shape evolution cannot
be deduced from the two-dimensional representation in the top panels of 
Fig.~\ref{fig:landscapes}, we provide a more intuitive picture of the elongation
of the nucleus through the contour plots of the density of AS prolate 
configurations of $^{240}$Pu in the bottom panel of the figure.
In all panels the region at large $\beta_{20}$ and $\beta_{22}$ is somewhat 
irregular: here (at least) two different valleys become close in energy. Since 
they are located at high excitation energy, we made no effort to completely 
resolve them.

\begin{figure}
\centering
\includegraphics[width=.4\textwidth]{{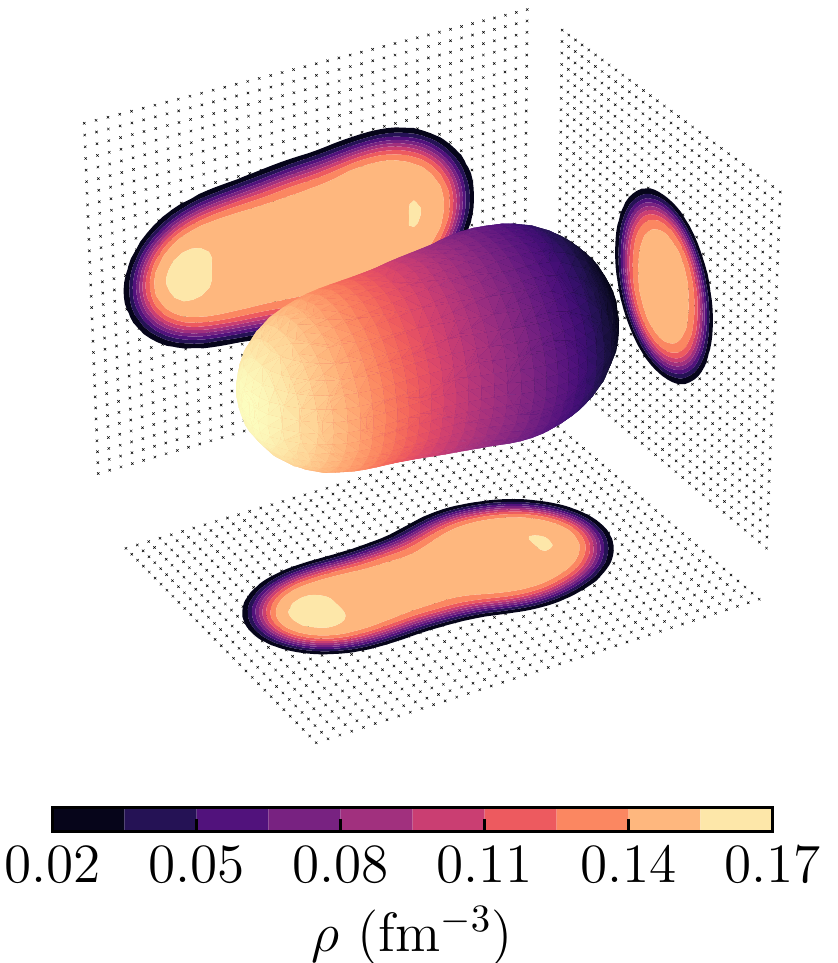}}
\caption{ Isodensity ($\rho = 0.02$ fm$^{-3}$) surface of $^{240}$Pu at 
          $\beta_{20} = 1.3$, $\beta_{22} = 0.03$, indicated by the red
          star in Fig.~\ref{fig:landscapes}. Two-dimensional contour plots of 
          the total density at $x=0$, $y=0$ and $z=0$ are also shown. 
          The colors of the 3D figure serve only to emphasize the 
          reflection asymmetry of the shape; the color bar refers exclusively 
          to the 2D projections.
          } 
\label{fig:fancy_3D}
\end{figure}

If one restricts the nucleus to axial symmetry, the corresponding least-energy 
fission path lies along the horizontal axis in Fig.~\ref{fig:landscapes}. We 
will call this path the axially symmetric path (ASP) in what follows. The two
saddle points along this path are the tops of the inner and outer peak, which 
respectively can reach heights of slightly more than 8 MeV and up to 7 MeV. 
When allowing for triaxial deformation however, the LEP in the full collective
space can detour along both peaks: compared to the ASP, the inner saddle point 
is lowered by more than an MeV while the effect for the outer saddle point 
accounts for a few hundred keV. These detours do not take the nuclei very
far from the ASP: $\beta_{22}$ remains small compared to $\beta_{20}$ and 
$\gamma$ remains below $15^{\circ}$ in the vicinity of the inner peak and even 
below $5^{\circ}$ beyond $\beta_{20} = 1$. Significant parts of the LEP 
overlap with the ASP: near the ground state, the superdeformed minimum and 
beyond the outer saddle point the nucleus prefers AS shapes. As an illustration 
of the triaxial and reflection-asymmetric configurations encountered near the 
outer saddle point, we show a three-dimensional isodensity surface for 
$^{240}$Pu in Fig.~\ref{fig:fancy_3D}. To underline the three-dimensional 
nature of this shape, we also provide two-dimensional contour plots of the 
density in each direction: no two of them are identical. 

The nuclei $^{236}$Np and $^{249}$Cm in Fig.~\ref{fig:landscapes} serve as
illustrations of the PES of odd-odd and odd-mass nuclei, respectively. They
do no appear qualitatively different from those of the even-even nuclei:
we have been able to properly converge our self-consistent blocked calculations 
in all regions of the collective space shown, yet we remind the reader that 
the evolution of the blocked quasiparticles along the PES can be quite
complicated.

While the general features of the PESes are the same for all five nuclei, 
differences are clearly visible. First, the deformations of ground state and 
fission isomers grow with increasing neutron number. Second, the height of the 
inner peak increases with neutron number: from about $6$ MeV for $^{232}$U to 
just over $8.8$ MeV for $^{249}$Cm.
The outer peak varies by less: $5.8$ MeV for $^{232}$U to $5.1$ MeV for $^{249}$Cm.
While the $(\beta_{20}, \beta_{22})$
coordinates of the LEPs in Fig.~\ref{fig:landscapes} look qualitatively 
similar in all cases, the impact of triaxial deformation on the total 
energy along these paths varies strongly with $N$: for $^{232}$U, triaxial 
deformation lowers the inner barrier by about an MeV while the effect for $^{249}$Cm 
is close to four MeV. Triaxial deformation also affects the outer barrier 
more strongly as $N$ increases, although this is not easily visible on 
Fig.~\ref{fig:landscapes}. The impact of triaxial deformation on the barriers is
sufficiently large to change their evolution with $N$: among these five isotopes, 
$^{249}$Cm has the largest inner barrier among these five nuclei when 
restricted to axial symmetry, but if we include $\beta_{22}$ as collective 
coordinate its barrier becomes the lowest one.

\begin{figure}
\centering
\includegraphics[width=.4\textwidth]{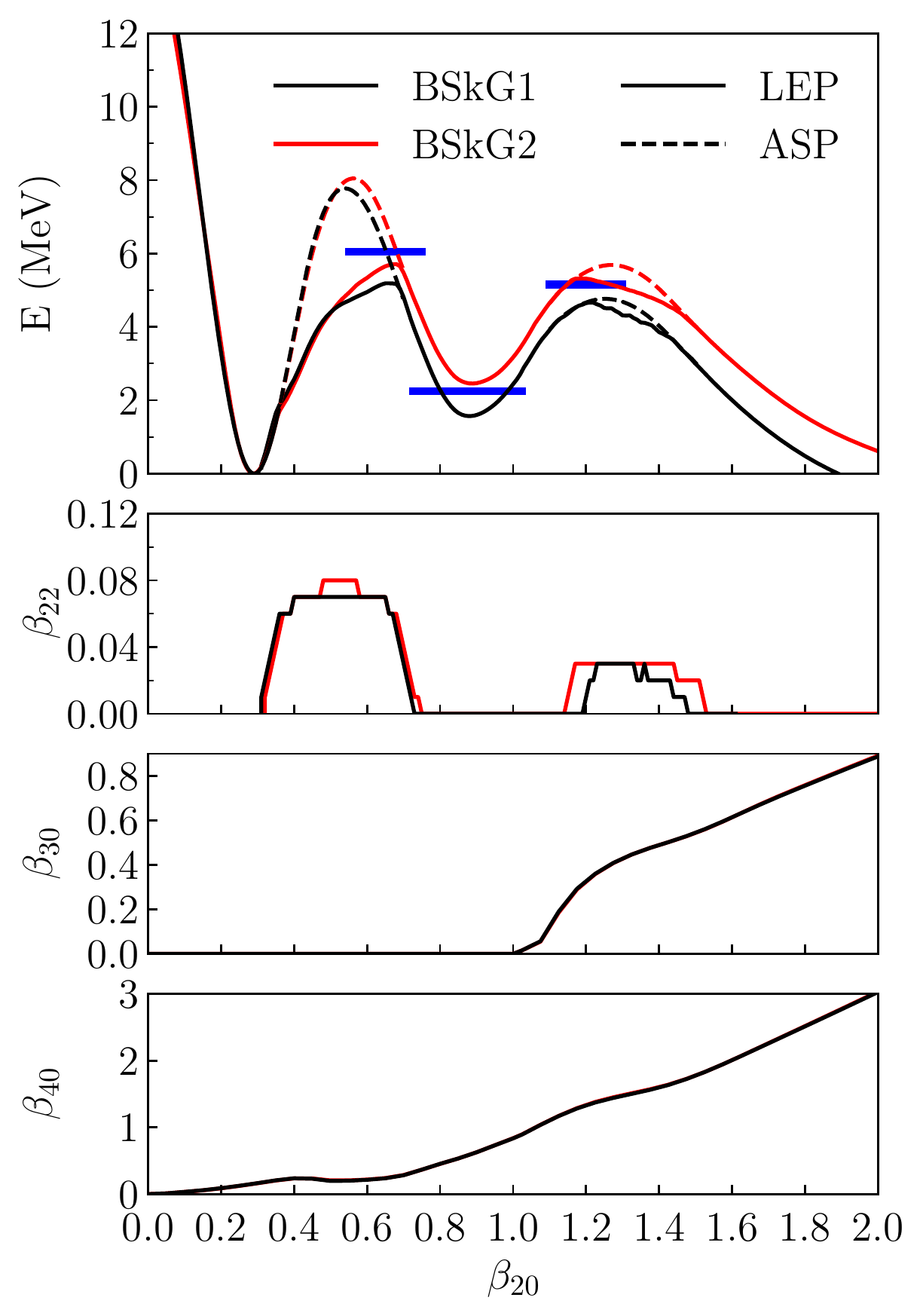}
\caption{ (Color online) 
          Top panel: Energy of $^{240}$Pu along the LEP (solid lines) and 
          ASP (dashed lines) as a function of $\beta_{20}$, 
          obtained with BSkG1 (black lines) and BSkG2
          (red lines), normalized to the g.s. energies obtained with 
          the respective models. We also show empirical 
          values for the primary and secondary barrier, as well as the isomer
          excitation energy, drawn at illustrative ranges of $\beta_{20}$ (blue lines). 
          Second, third and fourth panels: 
          $\beta_{22}$, $\beta_{30}$ and $\beta_{40}$ along the BSkG1 and BSkG2 
          LEPs. The octupole and hexadecapole moments of the nucleus along the 
          ASPs are indistinguishable from those along the LEPs and are not shown.
          }          
\label{fig:1D_Pu240}
\end{figure}

We will study barriers, isomer excitation energies and their evolution with 
particle number in a more systematic fashion in Sec.~\ref{sec:triax}.
For further illustrations, we restrict our attention for now to $^{240}$Pu and 
study its ASP and LEP in more detail in Fig.~\ref{fig:1D_Pu240}; the top 
panel shows the evolution of the total energy along the LEPs and ASPs (full
and dashed lines, respectively) obtained with BSkG1 and BSkG2 (black and red
lines, respectively), normalized to their respective g.s. minima. The 
RIPL-3 empirical values for the barrier heights and the isomer excitation 
energy from Tab.~\ref{tab:fission_fit_data}~\cite{Capote09,Samyn04} are shown 
by blue lines at illustrative ranges of $\beta_{20}$. The impact of triaxiality 
on both barriers can clearly be seen: for BSkG1, including this 
degree of freedom lowers the inner and outer barrier by about 2.6 MeV and 
about 120 keV respectively. For BSkG2, the effect on the inner barrier is 
smaller while that on the outer barrier is larger: they get lowered by about 
2.3 MeV and 350 keV, respectively. The impact of triaxial deformation we obtain
here is somewhat larger than that reported for the SLy5sX-family of Skyrme 
parameterizations in Ref.~\cite{Ryssens19}: for these, triaxial deformation
lowers the inner and outer barrier of $^{240}$Pu by about 1.5 MeV and 300 keV, 
respectively. The barriers and isomer excitation energy obtained with 
BSkG1 along the LEP match the reference values of Tab.~\ref{tab:fission_fit_data} 
within about an MeV. Although this level of agreement is already quite good, the 
BSkG2 results are much closer to the empirical values.

The other panels of Fig.~\ref{fig:1D_Pu240} show multipole moments other than 
$\beta_{20}$ along the LEPs obtained: from top to bottom $\beta_{22}$,
$\beta_{30}$ and $\beta_{40}$. While we do not show multipole moments such as 
$\beta_{32}$ or $\beta_{60}$, we remind the reader that these do not vanish 
and are naturally included in the self-consistent solution procedure. We take 
the smooth evolution of multipole moments, both those included in 
Fig.~\ref{fig:1D_Pu240} and those that are not, as an indication that the 
LEPs constructed present continuous trajectories. 
The sequence of shapes of the nuclear density along the path is robust, 
at least for this nucleus: $\beta_{30}$ and $\beta_{40}$ do not change when 
either exchanging BSkG1 for BSkG2 or the ASP for the LEP. Octupole 
deformation vanishes at small to moderate deformations and only beyond $\beta_{20} \sim 1.0$ 
does the LEP explore reflection-asymmetric shapes. The second quadrupole moment
$\beta_{22}$ remains small along the LEP for both models when compared 
to $\beta_{20}$. While the excitation energies of the saddle points along the
fission path are lowered significantly by the inclusion of $\beta_{22}$ as
collective coordinate, the shape evolution of the nucleus as measured by any 
other multipole moment is essentially unmodified. 

In passing, we note that both BSkG1 and BSkG2 reproduce the experimental information 
on the quadrupole and hexadecupole deformation of the g.s. of $^{240}$Pu. 
We calculate $\beta_{20,c} = 0.287, \beta_{40,c} \sim 0.15$ with no difference
between models and in excellent agreement with $\beta_{20,c} = 0.293(2), \beta_{40,c} = 0.16(4)$
of Ref.~\cite{Bemis73} and  $\beta_{20,c} = 0.292(2), \beta_{40,c} = 0.151(8)$
of Ref.~\cite{Zumbro86} as determined from Coulomb excitation and the analysis of muonic
$X$ rays, respectively\footnote{
We note that the tables of Ref.~\cite{Bemis73} and \cite{Zumbro86} report on 
deformation parameters that characterize the \emph{surface} of the 
nuclear shape. These are not equal to our $\beta_{\ell m,c}$, which characterize 
the shape of the nuclear \emph{volume}, see the discussion in 
Refs.~\cite{Ryssens19,Ryssens23} and references therein. For this reason, the 
experimental values for the $\beta_{\ell m,c}$ quoted in the text are charge multipole deformations 
that we consistently calculated from the electric transition moments~\cite{RingSchuck}
that are also provided by these references.
}.

\begin{figure}
\centering
\includegraphics[width=.4\textwidth]{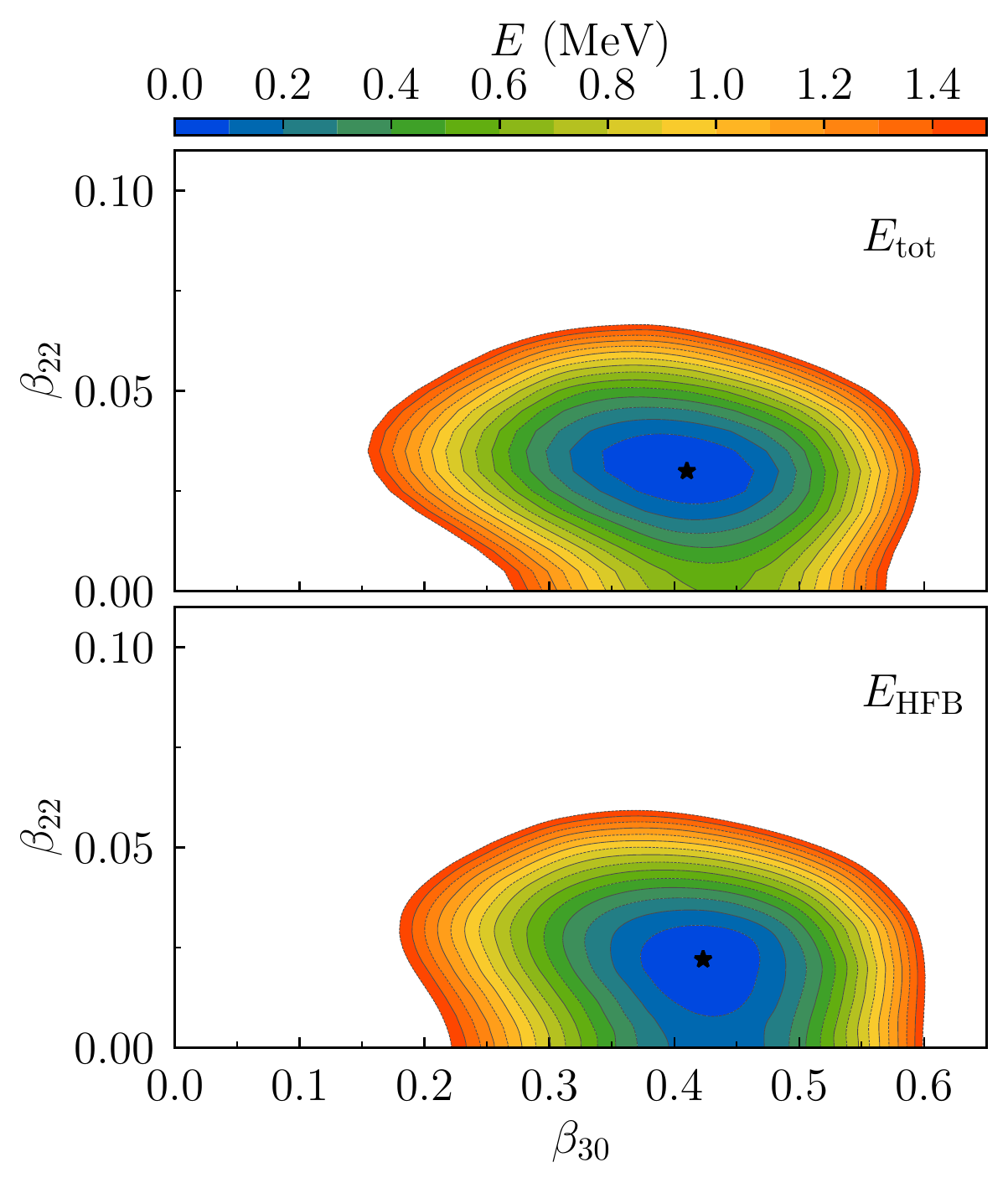}
\caption{ (Color online) 
         The total energy $E_{\rm tot}$ (top panel)  
         and the mean-field energy $E_{\rm HFB}$ (bottom panel)
         of $^{240}$Pu at (fixed) elongation $\beta_{20} = 1.30$, 
         normalized to their respective minima, as a function of $\beta_{22}$ and 
         $\beta_{30}$. Respective minima are indicated by black stars and 
         contour lines are 100 keV apart. }
\label{fig:top_of_the_barrier}
\end{figure}

Next, we illustrate the effect of octupole deformation on the PES and the 
fission path at large deformation. Fig.~\ref{fig:top_of_the_barrier} shows 
the total energy (top panel) and the mean-field energy (bottom panel) of 
$^{240}$Pu obtained with BSkG2 as a function of $\beta_{22}$ and $\beta_{30}$ 
at a fixed elongation $\beta_{20} = 1.3$. The minima of both surfaces are 
situated at almost identical octupole deformation, leading us to two conclusions.
First, a hypothetical LEP constructed from a 
three-dimensional PES using $\beta_{30}$ as additional collective coordinate 
would have been close to the one we obtain from our two-dimensional calculations, 
as would the barriers deduced from it. Second, Fig.~\ref{fig:top_of_the_barrier}
illustrates that the preference of the LEP for triaxial deformation is not 
\emph{only} due to the presence of the collective correction. For this
nucleus, the minimum of the uncorrected mean-field energy is not axially 
symmetric. The inclusion of $E_{\rm corr}$ moves the LEP to a slightly larger value of $\beta_{22}$
and enlarges the energy difference between triaxial and AS 
configurations, stabilizing the triaxial deformation when the 
energy surface is soft in $\beta_{22}$.

\begin{figure}
\centering
\includegraphics[width=.4\textwidth]{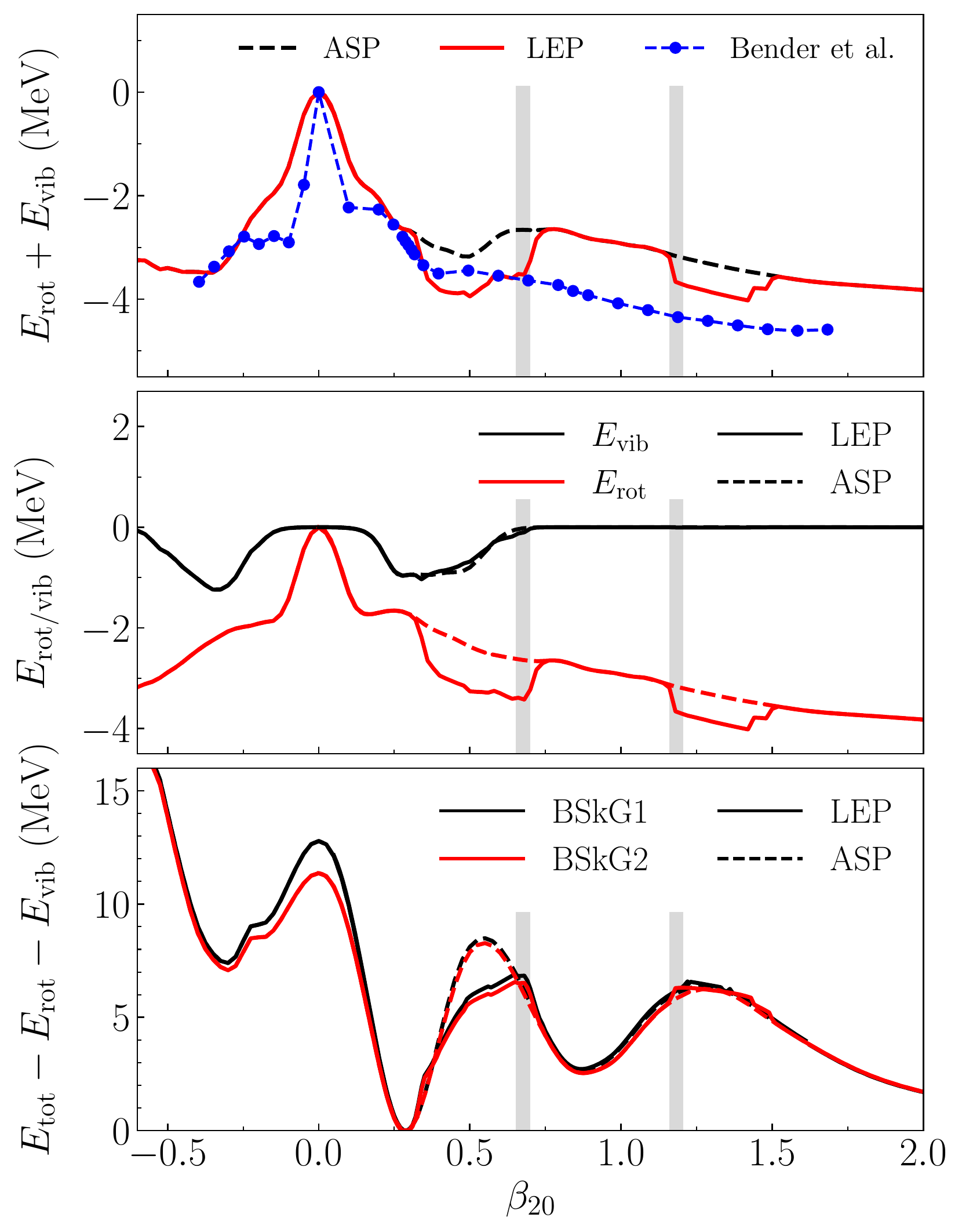}
\caption{ (Color online) 
          Illustration of the effect of the rotational and vibrational
          correction for $^{240}$Pu. 
          Top panel: the total collective correction for BSkG2 along the 
          LEP (red full line) and along the ASP (black dashed line), 
          compared to the correlation energy obtained by projection on angular momentum $J=0$ in Ref.~\cite{Bender04}, 
          using the SLy6 parametrisation of Skyrme's EDF (blue circles).
          Middle panel: vibrational energy (black curves) and rotational
          energy (red curves) for BSkG2 along the LEP (full lines) and the 
          ASP (dashed lines). Bottom panel: total energy minus the vibrational
          and rotational energy for BSkG1 (black curves) and BSkG2 (red curves)
          along the respective LEPs (full lines) and ASPs (dashed lines).   
          Note that the LEP is obtained from the \emph{total energy}.
          Faint grey bars are centered on the $\beta_{20}$ value of the inner and outer 
          saddle points along the LEP with a width of $\delta \beta_{20} = \pm 0.02$.
          }
\label{fig:collcorr}
\end{figure}

To illustrate the influence of the collective correction on the PES, we show 
its evolution along the ASP and LEP in the top panel of Fig.~\ref{fig:collcorr} 
for the BSkG2 model. The contributions of the vibrational and rotational 
correction along each path are plotted separately in the middle panel. The rotational correction 
contributes several MeV to the binding energy, except near the spherical point.
Along the ASP the size of this correction increases smoothly with deformation;
where the LEP passes through finite values of $\beta_{22}$ there is an extra
contribution on the order of 1 MeV that is associated with the extra
rotational degree of freedom. Along both paths, the rotational correction 
systematically \emph{decreases} both the barriers and the isomer excitation energies.
In general, this correction affects more the features of the PES at large deformation.
The vibrational correction on the other hand only contributes to the energy for modest deformations 
$0.1 \lesssim \beta_{20} \lesssim 0.55$: it vanishes near the isomer and the 
outer saddle point and contributes less than 100 keV near the inner saddle point 
on the LEP.  Because this correction increases the binding energy of the g.s., 
its presence indirectly \emph{raises} both barriers and the isomer excitation energy in a uniform way.

For comparison with Fig.~\ref{fig:1D_Pu240} we show the total energy without 
both corrections in the bottom panel of Fig.~\ref{fig:collcorr} for both BSkG1 and BSkG2: 
except at small deformation, the evolution of the energy with deformation for 
both models is nearly identical\footnote{The total energy, on which the 
construction of the LEP is based, is a smooth function as seen on the top panel 
of Fig.~\ref{fig:1D_Pu240}. Its decomposition is not: the discrete steps in 
$\beta_{22}$ of the LEP are visible in the 
non-smooth parts of the LEP curves in Fig.~\ref{fig:collcorr}.}.
This is supported by Hartree-Fock calculations of semi-infinite 
nuclear matter along the lines of Ref.~\cite{Jodon16}: for both models we obtain 
essentially identical values of the surface energy coefficient
$a^{(\rm HF)}_{\rm surf} = 17.9$ MeV. This is somewhat larger than the value 
for the SLy5s1 parameterization, $a^{(\rm HF)}_{\rm surf} = 17.55$ MeV, which 
was found to be close to optimal for the description of fission properties 
in Ref.~\cite{Ryssens19} when not including a rotational correction. 
This difference illustrates the impossibility to establish a unique 
model-independent empirical value for $a^{(\rm HF)}_{\rm surf}$: the rotational 
correction characterises the many-body state and grows with deformation and 
thereby contributes to the deformation energy. But the rotational correction 
does not contribute to the surface energy coefficient $a^{(\rm HF)}_{\rm surf}$,
which is a state-independent characteristic of the effective interaction that 
quantifies the energy loss from the presence of a surface. Hence, when 
fission properties are included in the parameter fit, models without quantal 
corrections will have a lower surface energy than models that include such corrections.

The total collective correction we employ here qualitatively agrees with the 
correlation energy obtained through rotational symmetry restoration for 
AS configurations of $^{240}$Pu using the SLy4 parameterization in 
Ref.~\cite{Bender04}, shown by blue circles in the top panel of Fig.~\ref{fig:collcorr}. 
The comparison is far from perfect, particularly at large deformation, but firm conclusions
are hard to draw due to the differences between conditions of the calculations. 
In any case, the phenomenological treatment of collective motion remains a 
weak point of our approach that will require improvement in the future.

%%%%%%%%%%%%%%%%%%%%%%%%%%%%%%%%%%%%%%%%%%%%%%%%%%%%%%%%%%%%%%%%%%%%%%%%%%%%%%%%
\section{Actinide fission barriers with BSkG2}
\label{sec:systematics}

\subsection{ Global description of barriers and isomers}
\label{sec:fission_global}

In this section we embark on a more systematic description of the fission properties 
of $45$ $Z \geq 90$ nuclei in the RIPL-3 database: 14 are even-even and 8 are
odd-odd, while among the 23 odd-mass systems there are 6 odd-$Z$ and 17 odd-$N$ 
nuclei. For 28 nuclei among this set\footnote{Ref.~\cite{Samyn04} lists in 
fact 30 isomer excitation energies, but we drop for simplicity the values for
$^{235}$Pu and $^{244}$Bk for which RIPL-3 lists no empirical barriers.}, 
Ref.~\cite{Samyn04} lists a value for the excitation 
energy of the isomer: 8 are for even-even nuclei, 4 for odd-odd nuclei and 
16 concern odd-mass isotopes. Of the latter, 4 and 12 values are given for
odd-$Z$ and odd-$N$ isotopes.  

To quantify the performance of BSkG1 and BSkG2 as well as other models in the 
literature, we will investigate their rms and mean deviations, $\sigma(O)$ and $\bar{\epsilon}(O)$,
for three quantities $O = E_{\rm I}$, $E_{\rm II}$ and $E_{\rm iso}$, that are 
the primary and secondary barriers and the excitation energy of the fission 
isomer as introduced in Sec.~\ref{sect:parameter:adjustment}. To avoid confusion 
about signs, we define the mean deviation of a quantity for which we have access
to $M$ empirical and calculated values as:
\begin{align}
\bar{\epsilon}(O) &= \frac{1}{M}\sum_{i=1}^{M}       \left(O_{i}^{\rm emp} - O_{i}^{\rm th}\right)
\label{eq:meandeviation} \, .
\end{align}
Using this convention, positive (negative) values of $\bar{\epsilon}(O)$ mean that 
our calculations underestimate (overestimate) the empirical values.

\begin{table}[]
\centering
\begin{tabular}{llrlllll}
\hline 
\hline
          & & & \multicolumn{2}{c}{BSkG1} && \multicolumn{2}{c}{BSkG2}     \\ 
\noalign{\smallskip}
\cline{4-5}\cline{7-8}
\noalign{\smallskip}
     &  & $M$ & \multicolumn{1}{c}{$\sigma$} & \multicolumn{1}{c}{$\bar{\epsilon}$}
           && \multicolumn{1}{c}{$\sigma$} & \multicolumn{1}{c}{$\bar{\epsilon}$} \\
\hline
Even-even & $E_{\rm I}  $  &14 & 0.94  & $+0.90$ && 0.45 & $+0.31$\\
          & $E_{\rm II} $  &14 & 0.83  & $+0.67$ && 0.46 & $+0.01$\\
          & $E_{\rm iso}$  & 8 & 0.63  & $+0.52$ && 0.53 & $-0.38$\\
\noalign{\smallskip}
Odd-$Z$     & $E_{\rm I}$    & 6 & 0.66  & $+0.52$ && 0.41 & $-0.03$\\
          & $E_{\rm II}$   & 6 & 0.69  & $+0.62$ && 0.28 & $+0.10$\\
          & $E_{\rm iso}$  & 4 & 0.66  & $+0.62$ && 0.40 & $-0.29$\\  
\noalign{\smallskip}
Odd-$N$     & $E_{\rm I}$    &17 & 0.96  & $+0.87$ && 0.5  & $+0.34$ \\
          & $E_{\rm II}$   &17 & 0.93  & $+0.72$ && 0.55 & $+0.12$ \\
          & $E_{\rm iso}$  &12 & 1.35  & $+0.85$ && 0.46 & $-0.35$ \\  
\noalign{\smallskip}
Odd-odd   & $E_{\rm I}$    & 8 & 0.73  & $+0.67$ && 0.28 & $+0.13$\\
          & $E_{\rm II}$   & 8 & 0.95  & $+0.84$ && 0.43 & $+0.24$\\
          & $E_{\rm iso}$  & 4 & 0.62  & $+0.50$ && 0.57 & $-0.43$\\  
\noalign{\smallskip}
  Total   & $E_{\rm I}$    &45 & 0.88  & $+0.80$ && 0.44 & $+0.24$\\
          & $E_{\rm II}$   &45 & 0.87  & $+0.71$ && 0.47 & $+0.10$ \\
          & $E_{\rm iso}$  &28 & 1.00  & $+0.67$ && 0.49 & $-0.36$\\  
\hline
\hline
\end{tabular}
\caption{Rms $\sigma(O)$ and mean deviations $\bar{\epsilon}(O)$ 
          of the BSkG1 and BSkG2 models, with respect to RIPL-3 reference values 
          for the primary and secondary barriers~\cite{Capote09}          
          and isomer excitation energies from Ref.~\cite{Samyn04} for different 
          subsets of nuclei: even-even nuclei, odd-mass nuclei with odd $Z$, odd-mass
          nuclei with odd $N$ and odd-odd nuclei. $M$ indicates the number of empirical 
          values available for each subset.
          All energies are expressed in MeV. 
          }
\label{tab:barrier_properties}
\end{table}

\begin{figure}
\centering
\includegraphics[width=.475\textwidth]{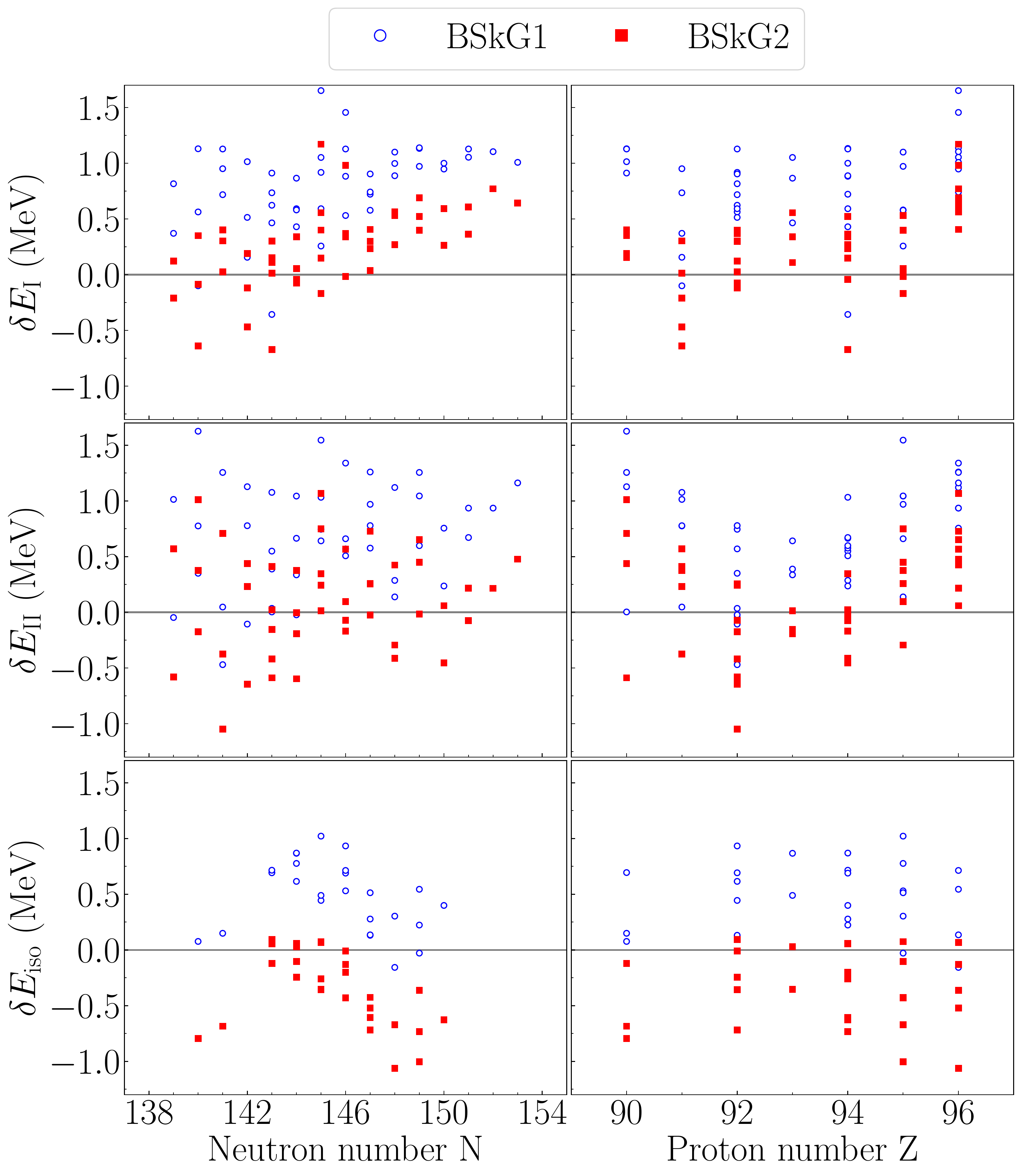}
\caption{ (Color online)
          Difference between calculated and reference values for the primary 
          barrier heights (top panel), secondary barrier heights
          (middle) and isomer excitation \textit{energies} (bottom panel), using BSkG1 
          (blue open circles) and BSkG2 (red filled squares). Positive (negative) 
          values for all three differences mean that calculated results are
          smaller (larger) than the reference values.
         }
\label{fig:barriers}
\end{figure}

The values of these deviations for BSkG2 are given in Table~\ref{tab:barrier_properties}, 
calculated for the complete set of nuclei, but also for subsets of nuclei
separated by the number parity of the two nucleon species. For comparison, we also list the values 
for the BSkG1 model obtained through identical calculations but for the useage 
of the Equal-Filling Approximation (EFA)~\cite{Perez08} for odd-mass and odd-odd nuclei.
We show in Fig.~\ref{fig:barriers} the differences for individual nuclei between the 
empirical values and the calculations, $\delta O \equiv O^{\rm emp} - O^{\rm th}$, 
as a function of neutron (left column) and proton number (right column).

BSkG1 reproduces the empirical values for barriers and isomer excitation 
energies roughly within 1.5 MeV; the largest deviation between empirical 
and calculated values for primary barries occurs for $^{241}$Cm at roughly $1.6$ MeV.
The total rms deviations for the three quantities for BSkG1 are all slightly below 1 MeV. This level of agreement 
with the reference values is not typical among all Skyrme parameterizations, 
many of which overestimate fission barriers by as much as 10 MeV~\cite{Jodon16}. 
As we will see in Sec.~\ref{sec:comparison}, it is however roughly 
representative of the subset of models whose parameter adjustment considered 
fission in one way or another~\cite{Kortelainen14,Bartel82}. Since the 
construction of BSkG1 did not involve fission properties, its quality 
in this respect is thus somewhat remarkable. Nevertheless, the mean deviations 
of this model are all larger than 0.6~MeV, indicating a systematic 
underestimation of inner and outer barriers as well as isomer excitation energies. 

By adding the vibrational correction and adjusting to fission data, BSkG2 does
significantly better than BSkG1 for the primary barriers. The total BSkG2 
rms deviation for this quantity is only $\sigma(E_{\rm I}) = 0.44$ MeV, 
which is half that of BSkG1. The BSkG2 mean deviation of the primary 
fission barrier is much smaller, $\bar{\epsilon}(E_I) = +0.24$ MeV, but remains 
non-zero and positive. The largest difference between the empirical values and the
calculations still occurs for $^{241}$Cm, but this is the only nucleus whose
primary barrier is not reproduced within one MeV. The reproduction of the 
secondary barriers is also much improved, as shown by an rms deviation of 
$\sigma(E_{\rm II}) = 0.47$ MeV. The corresponding mean deviation, 
$\bar{\epsilon}(E_{\rm II}) = +0.10$ MeV, is even smaller than that of the 
primary barriers but also indicates a small but systematic underestimation 
of the secondary barriers. The BSkG2 rms deviation of isomer excitation energies is less than half of its predecessor, 
$\sigma(E_{\rm iso}) = 0.49$ MeV, which is as accurate as the description
of the barriers. However, the model overestimates $E_{\rm iso}$ for nearly all 
nuclei we consider, which is reflected in the large \emph{negative} mean 
deviation, $\bar{\epsilon}(E_{\rm iso}) = -0.36$ MeV, in contrast to the 
systematic underestimation of this quantity by BSkG1. The mean and rms 
deviations restricted to subsets of nuclei show that these observations generally 
hold separately among the subsets of even-even, odd-$Z$, odd-$N$ and odd-odd
nuclei. One difference among subsets concerns the mean deviation of the 
primary barrier: for the small number of odd-Z and odd-odd systems it is smaller 
($\bar{\epsilon}(E_{\rm I}) = -0.03$ and $+0.13$, respectively) than that for 
the large number of even-even and odd-odd nuclei 
($\bar{\epsilon}(E_{\rm I}) = +0.31$ and $+0.34$, respectively). 
Despite this difference and other more minor ones, 
the deviations in Table~\ref{tab:barrier_properties} are comparable among 
all subsets, indicating that our choice to include only 12 even-even nuclei
in the objective function did not bias the parameter adjustment. 

The mean deviations of BSkG2 are small but non-zero: the tendency of the
model to overestimate isomer excitation energies while slightly underestimating 
barriers can be see in Fig.~\ref{fig:barriers}. This deficiency cannot be 
solved by further fine-tuning of the collective correction: as is clear from the 
discussion in Sec.~\ref{sec:Pu240}, the parameters of $E_{\rm corr}$ do not 
offer the possibility to simultaneously \emph{raise} the barriers and 
\emph{lower} the isomer excitation energies. It is tempting to look for further
trends in the deviations as a function of particle number in Fig.~\ref{fig:barriers}, 
but we remind the reader that (i) empirical values are only available for a
tiny fraction of the nuclei relevant to nucleosynthesis and (ii) that the 
empirical values themselves are subject to uncertainties, as we remarked
on in Sec.~\ref{sect:parameter:adjustment}.
Furthermore, the mean deviations of BSkG2 are not much larger than the 
numerical precision of our calculations, roughly 100 keV. One could argue that 
the BSkG2 mean deviation for the secondary barriers essentially vanishes, even 
though this quantity remains the average of much larger errors, see Fig.~\ref{fig:barriers}.
 
The accuracy of BSkG2 as reflected by the small rms and mean deviations in 
Table~\ref{tab:barrier_properties} is excellent compared to other large-scale 
models of fission properties in the literature and it is striking that the rms
deviation of the primary barriers, secondary barriers and isomer excitation 
energies are all comparable. We will discuss these observations in more detail 
in Sections~\ref{sec:triax} and \ref{sec:comparison}, but note here already
that the success of BSkG2 is not solely due to inclusion of the vibrational
correction: the first step of the parameter adjustment resulted in a 
parameterization with fission properties similar to those of BSkG1, even though 
no information on the properties of nuclei at large deformation entered this phase 
of the fit. The quality of this intermediate parameterization for fission is 
not a given; instead it results from the choices made in our modelling of the 
nuclear binding energy such as the inclusion of the two-body part of the 
centre-of-mass correction. The adjustment procedure of many available Skyrme 
parameterizations only accounted for its one-body part, resulting in significantly 
larger surface tensions~\cite{Bender00,daCosta22} and an overestimation of 
fission barriers by 10~MeV or more~\cite{Jodon16} unless explicitly constraining
deformation properties as done for SkM* \cite{Bartel82} or SLy5s1 \cite{Jodon16}.
Through fine-tuning of the parameters of the collective correction, we were 
able to reduce the rms and mean deviations of fission properties from values 
comparable to those of BSkG1 to the final BSkG2 values in Table~\ref{tab:barrier_properties}. 
We repeat here that the quality of the model for g.s. properties did not 
meaningfully suffer from this second step of the parameter adjustment: BSkG2 
achieves an rms deviation on essentially all known nuclear masses of 
AME2020~\cite{Wan21} of $\sigma(M) = 0.678$ MeV, slightly better than that of 
BSkG1 and competitive with most other models in the literature~\cite{Ryssens22}.

\begin{figure}
\centering
\includegraphics[width=.45\textwidth]{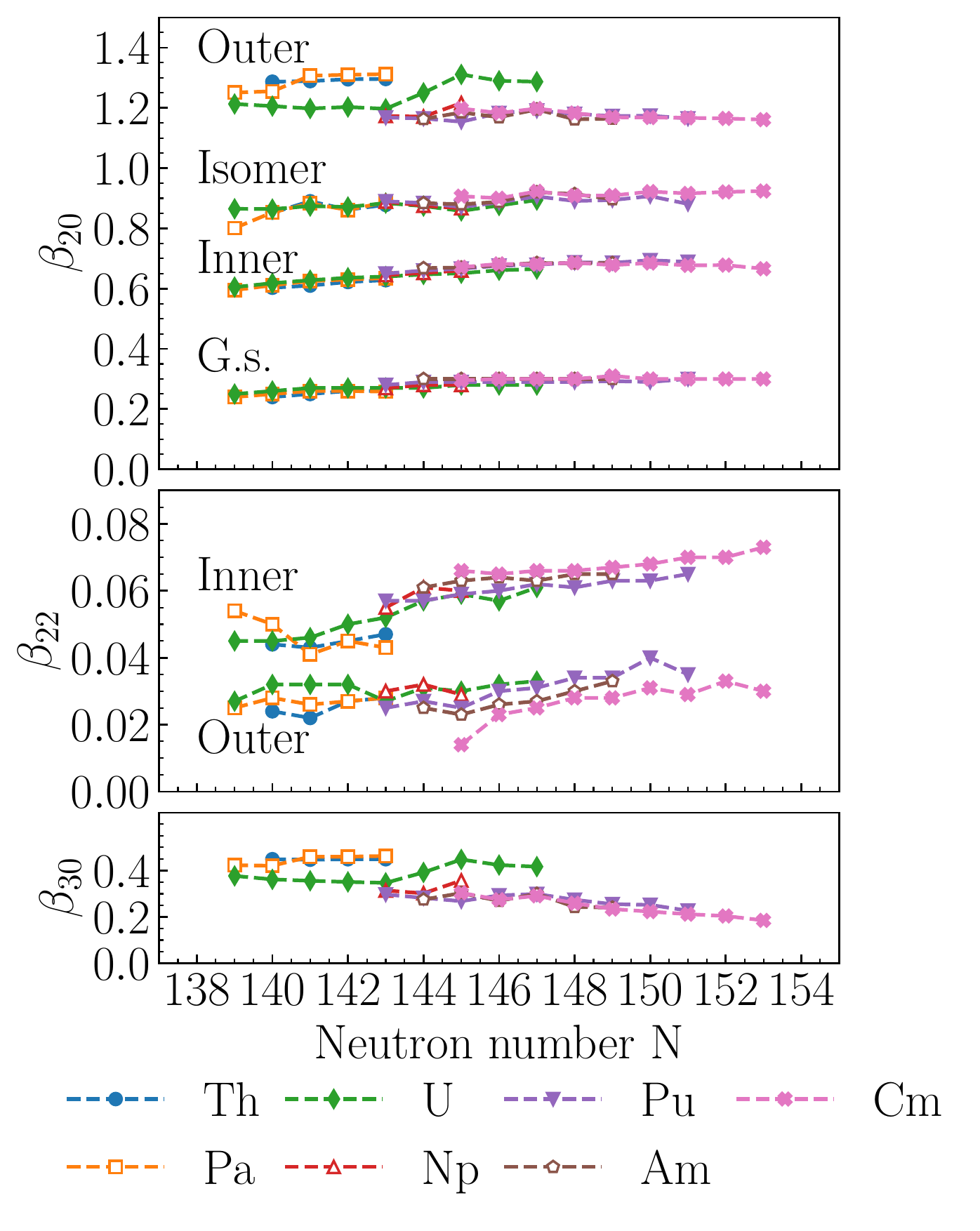}
\caption{ (Color online)
          Calculated quadrupole deformations $\beta_{20}$ (top panel) and $\beta_{22}$ (middle panel) 
          as well as octupole deformation ($\beta_{30}$, bottom panel) 
          for different features of the PES as a function of neutron number: 
          the g.s. minimum, the inner saddlepoint, the superdeformed isomeric
          minimum and the outer saddle point. Note that $\beta_{22} = \beta_{30} = 0$ for all 
          ground states and isomers and $\beta_{30} = 0$ in all cases save
          for the outer saddle points; these curves are not shown. 
          Even-$Z$ isotopes are drawn with full symbols, 
          odd-$Z$ with empty symbols.
          }
\label{fig:deformations}
\end{figure}

To conclude this section, we confirm that the PES of all nuclei we consider 
here has a topography similar to those discussed in Sec.~\ref{sec:Pu240}.
In Fig.~\ref{fig:deformations}, we illustrate this similarity by showing the 
location of the saddle points and isomers in terms of the quadrupole deformations 
$\beta_{20}$ and $\beta_{22}$, as well as the octupole deformation $\beta_{30}$.
The multipole moments of all three quantities evolve slowly and smoothly with 
neutron number. They depend only slightly on proton number, with the exception 
of the isotopes with lowest $Z$ (Th, Pa,U) whose outer barrier is somewhat 
different from the trend of the other isotopic chains. We point out that 
\emph{all} inner \emph{and} outer saddle points are located at 
non-zero values of $\beta_{22}$. Although the corresponding curves are not drawn in
Fig.~\ref{fig:deformations}, all ground state and isomeric minima are axial and
reflection-symmetric, meaning that for those features of the PESes $\beta_{22} = \beta_{30} = 0$. Although 
this cannot be deduced from the tables and figures shown so far, the 
superdeformed minimum is always higher in energy than the g.s. minimum near 
$\beta_{20} \sim 0.3$: $E_{\rm iso}$ is positive in all cases, even for BSkG1 
which systematically underestimates this quantity.

One minor further difference separates the Th isotopes from the other nuclei: 
their PESes show a shallow third minimum at $\beta_{20} \sim 1.6$ and a third,
barely pronounced, saddle point near $\beta_{20} \sim 2.0$ which is several MeV 
lower than even the secondary barrier and is thus of little consequence for
our discussion. This topography is similar to those described in 
Refs.~\cite{Berger89,Rutz95}, although in other calculations this outermost
(third) saddle point can determine the primary barrier~\cite{Lemaitre18}. 
Exploratory calculations show that Th isotopes with $N\leq 138$, isotopes
that do not figure in this study, can acquire an octupole deformation in their g.s.
when calculated with BSkG2.

%%%%%%%%%%%%%%%%%%%%%%%%%%%%%%%%%%%%%%%%%%%%%%%%%%%%%%%%%%%%%%%%%%%%%%%%%%%%%%%%
\subsection{The impact of triaxiality}
\label{sec:triax}

For the five examples of Sec.~\ref{sec:Pu240}, we found that the inclusion
of triaxial deformation has a large impact on inner barrier and a smaller 
effect on the outer barrier. In this section, we 
investigate this effect for the complete set of nuclei and study its 
evolution with particle number. This section does not discuss isomer excitation 
energies; the corresponding mean-field configurations are all axially symmetric.

\begin{figure}
\centering
\includegraphics[width=.45\textwidth]{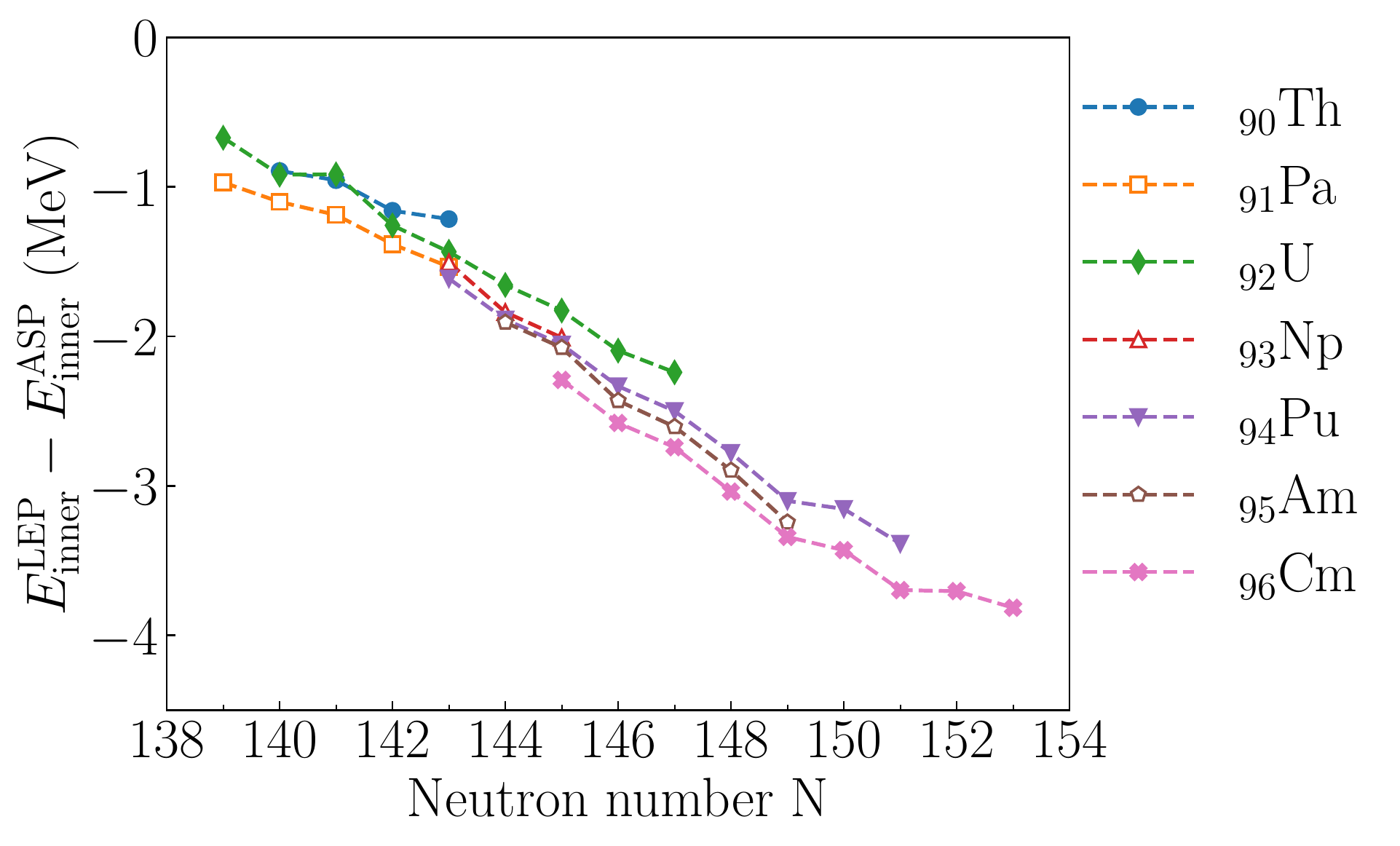}
\caption{(Color online) 
          Effect of triaxial deformation on the inner fission barrier 
          $E^{\rm LEP}_{\rm inner} - E^{\rm ASP}_{\rm inner}$ as a function of neutron 
          number $N$. Isotopic chains with even $Z$ and odd $Z$ are drawn with full 
          and empty symbols, respectively.}
\label{fig:effect_triax_inner}
\end{figure}

\begin{figure}
\centering
\includegraphics[width=.45\textwidth]{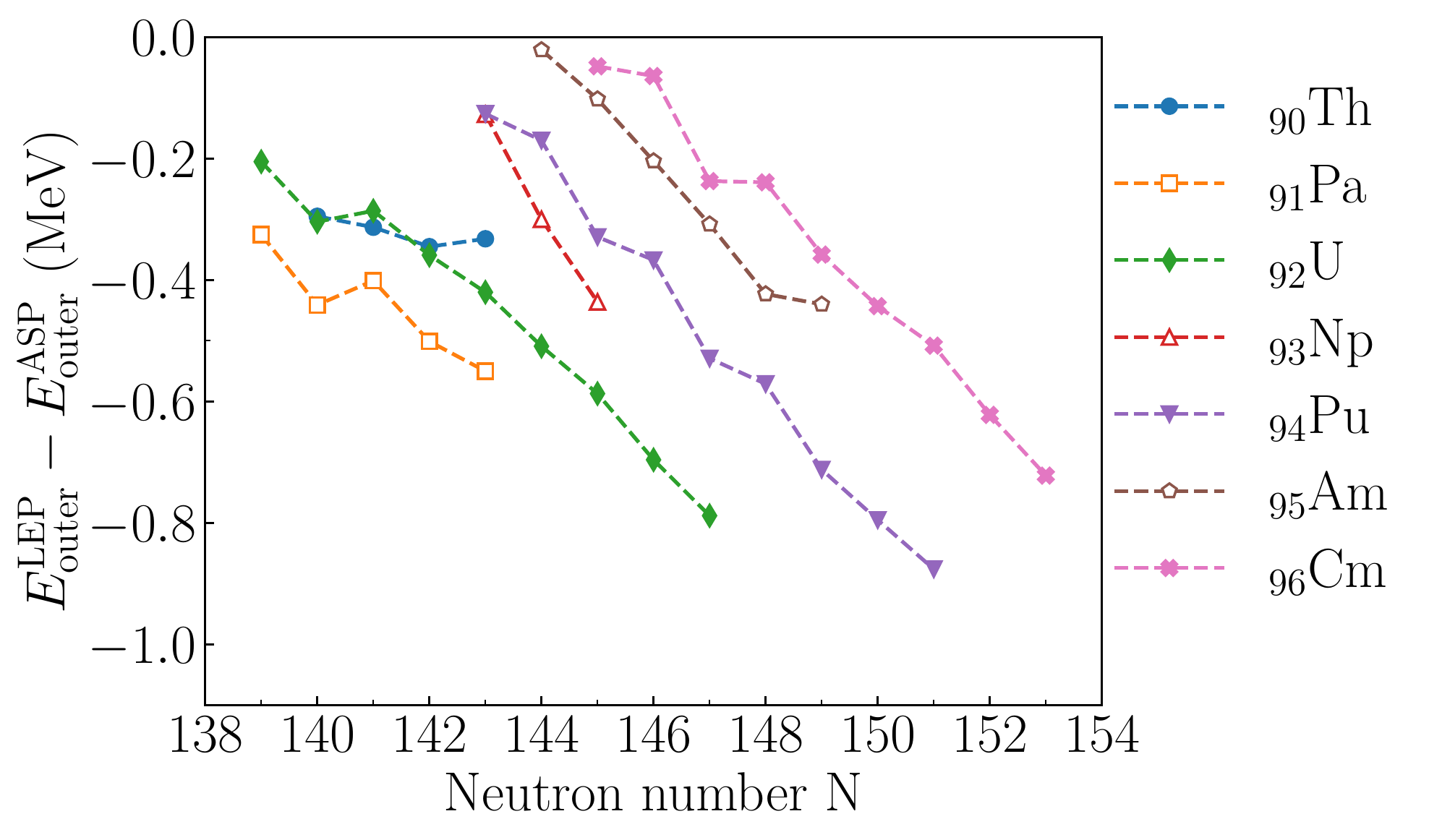}
\caption{(Color online) Same as Fig.~\ref{fig:effect_triax_inner}, but for the outer barrier.}
\label{fig:effect_triax_outer}
\end{figure}

We show in Fig.~\ref{fig:effect_triax_inner} the energy difference between
the inner barriers along the two fission paths: LEP$-$ASP. The difference 
of both barriers grows rapidly with increasing neutron number, ranging from 
about 650 keV for $^{231}$U to slightly less than 4~MeV ($^{249}$Cm). This 
is a significant correction for nuclei with $N\sim 140$ and is an enormous 
change for the nuclei with $N \sim 150$, for which the empirical primary 
barriers are on the order of 5~MeV. Similarly, 
Fig.~\ref{fig:effect_triax_outer} shows the 
energy difference of the outer barriers along the LEP and the ASP. The impact of 
triaxial deformation on the outer barrier is more modest, but is larger than 
400~keV for most of the nuclei we consider, ranging from virtually no effect for 
$^{240}$Am to about 850 keV for $^{247}$Pu. Although we do not show the corresponding
curves, the energy differences obtained with BSkG1 are very similar.

A striking aspect of Fig.~\ref{fig:effect_triax_inner} is its regularity: 
the triaxial energy gain for all isotopic chains fall almost on the same 
curve, with (i) only little deviation from a simple linear trend and (ii) 
very little dependence on the proton number. The latter observation suggests 
that the development of triaxial deformation near the inner barrier for
actinide nuclei is largely driven by the neutrons. It seems that the neutrons 
play a similar role near the outer barrier: within a given isotopic
chain, the energy difference increases linearly with neutron number. Along 
an isotonic chain however, the difference between LEP and ASP decreases with 
increasing $Z$ except for the Th isotopes. This suggests that triaxial 
deformation near the outer barrier is more of a competition between neutron 
and proton shell effects. 

\begin{figure}
\centering
\includegraphics[width=0.4\textwidth]{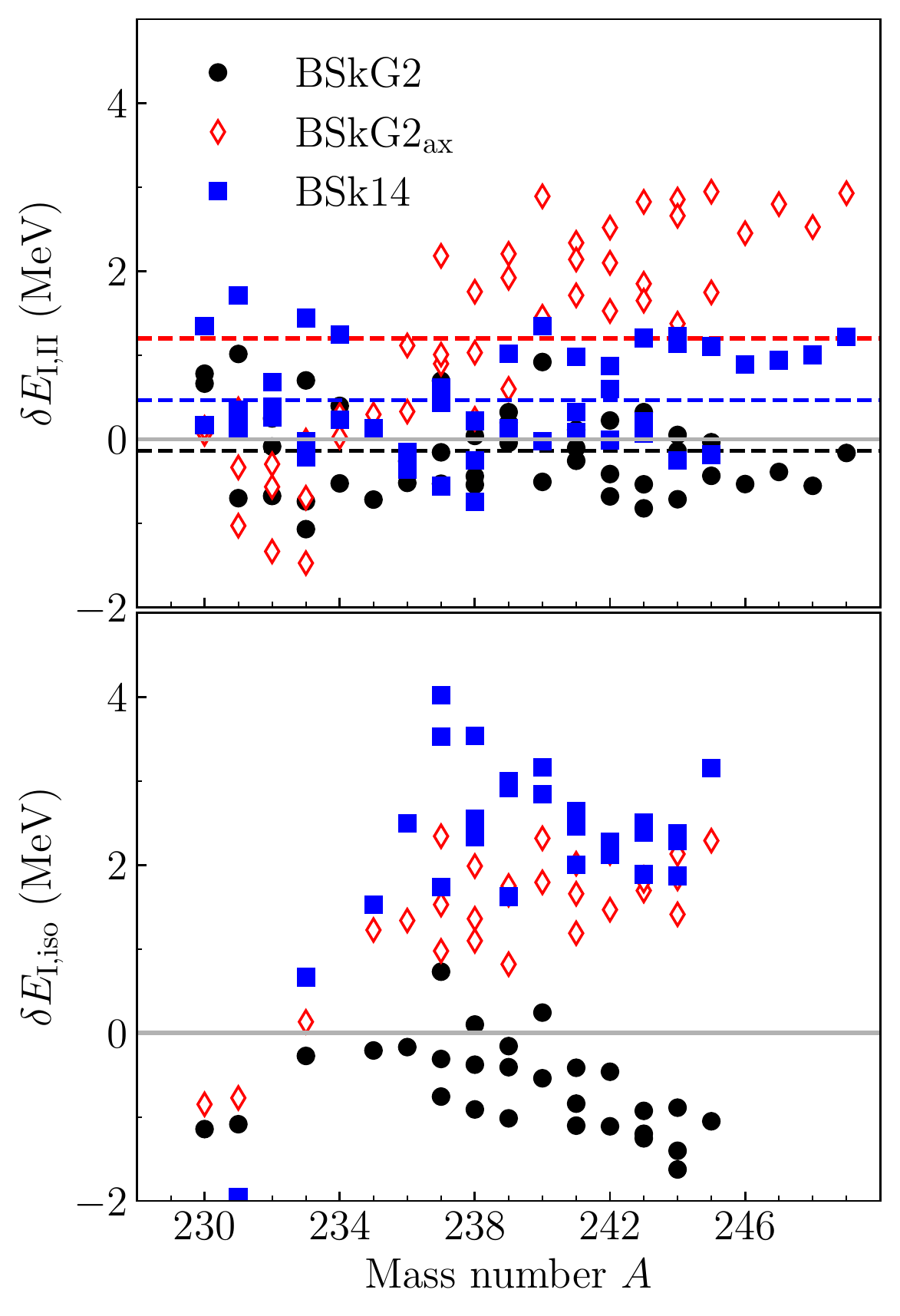}
\caption{Differences of fission properties $\delta E_{\rm I, II}$  (top panel)
         and $\delta E_{\rm I, iso}$ (bottom panel) for BSkG2 (black circles)
         and BSk14 (blue squares). We also show results constrained to axial
         symmetry for BSkG2 (empty red diamonds). The horizontal dashed 
         lines in the top panel indicate the average deviation in each case.}         
\label{fig:axial_diff}
\end{figure}

The effect of triaxial deformation is large, particularly for the inner barrier.
However, it is not a priori clear that including triaxial deformation necessarily 
leads to an improved quantitative description of fission properties in a 
framework like ours, which is equipped with a collective correction whose 
parameters are adjusted to fission data. To clearly understand 
the added value of including triaxial deformation at significant
computational cost, we study how well BSkG2 reproduces \emph{differences} of fission properties:
\begin{subequations}
\begin{align}
\label{eq:diff1}
\delta E_{\rm I, II}  & = E^{\rm calc}_{\rm I} - E^{\rm calc}_{\rm II}  - (E^{\rm emp}_{\rm I} - E^{\rm emp}_{\rm II})\, , \\
\delta E_{\rm I, iso} & = E^{\rm calc}_{\rm I} - E^{\rm calc}_{\rm iso} - (E^{\rm emp}_{\rm I} - E^{\rm emp}_{\rm iso})\, .
\label{eq:diff2}
\end{align} 
\end{subequations}
Since triaxial deformation impacts all three fission properties we consider 
differently, these differences emphasise the role of triaxiality. We show both 
$\delta E_{\rm I, II}$ and $\delta E_{\rm I, iso}$ in the two panels of 
Fig.~\ref{fig:axial_diff}: values obtained from a complete calculation with 
BSkG2 and those obtained from the ASP, labelled as BSkG2$_{\rm ax}$. The full 
BSkG2 calculation with triaxial deformation describes the difference of fission 
barriers with high accuracy but somewhat overestimates the difference 
between primary barrier and isomer excitation energies. Without triaxial 
deformation, the differences between (i) primary and secondary barrier and (ii) 
primary barrier and isomer excitation energy are enormously overestimated by 
the model and a spurious trend with mass number can be seen for the former. 

It is not possible to correct the deficiencies of the axially
symmetric calculation shown in Fig.~\ref{fig:axial_diff} by further adjustments
of the parameters of the collective correction. First off, the definitions in 
Eqs.~\eqref{eq:diff1} and \eqref{eq:diff2} essentially eliminate the 
influence of the vibrational correction which affects both barriers and the 
isomer in a roughly equal fashion. Second, changing 
the size of the rotational correction can alleviate the effect: for smaller 
values of $b$ the difference between primary and
secondary barrier will grow larger for the lighter nuclei and smaller for the 
heavier nuclei in Fig.~\ref{fig:axial_diff}.\footnote{An increase of the rotational
correction lowers the outer barrier more than the inner barrier, see the 
discussion around Fig.~\ref{fig:collcorr}. For the heavier nuclei, the inner one
tends to be the primary barrier such that an increase of $b$ implies a larger 
difference between primary and secondary barriers. For the lighter nuclei, 
the opposite happens.} 
To bring $\delta E_{\rm I, II}$ close to zero for the heavy nuclei however, 
one would have to resort to unreasonably small and possibly even negative values of $b$.

We cannot exclude that changing other aspects of the model (such as the 
surface energy) would not result in a better description of the difference of 
barriers in axially symmetric calculations. As illustration, we also include in 
Fig.~\ref{fig:axial_diff} results for BSk14: this model includes a 
rotational correction and vibrational correction adjusted in a similar fashion 
as those of BSkG2 but did not allow for triaxial deformation~\cite{Goriely07}\footnote{We also note that both models were
constructed with rather different numerical schemes: BSk14 relied on an 
expansion in a limited number of harmonic oscillator shells while we rely here
on a coordinate space representation.}. BSk14 does significantly better than 
the axial calculations with BSkG2 but still overestimates the difference 
between primary and secondary barrier, particularly for the lightest and 
heaviest nuclei. As we will see in Sec.~\ref{sec:comparison}, the systematic 
overestimation of $\delta E_{\rm I, II}$ is a generic feature of models that do not account 
for triaxial deformation. This is reflected in the average values of 
$\delta E_{\rm I,II}$: it exceeds 1~MeV for axial calculations with BSkG2 and is
about $+0.46$~MeV for BSk14, while the BSkG2 value of $-0.14$~MeV is a few times 
smaller.

The curves in Figs.~\ref{fig:effect_triax_inner} and~\ref{fig:effect_triax_outer} 
suggest that the effect of triaxial deformation grows even larger for more 
neutron-rich nuclei for which empirical values are 
not available. Since the appearance of triaxial deformation is linked to shell effects, 
it is unlikely that the effect of triaxial deformation across larger ranges of 
neutron and proton numbers will be as regular. Nevertheless, the role of 
triaxial deformation is not unique to the actinide region: for many superheavy 
nuclei it is found that triaxiality lowers the fission barrier by several MeV 
in both EDF-based and non-self-consistent models~\cite{Cwiok96,Bender98,Kowal10,Jachim17,Ryssens19b}.
We are currently extending the calculations presented here to thousands of 
heavy and superheavy nuclei far from stability, as a significant lowering of 
their fission barriers compared to earlier calculations has the potential 
to strongly modify the role of fission in r-process nucleosynthesis.

Triaxial deformation is somewhat routinely accounted for in EDF-based studies 
of the inner barrier of limited numbers of actinide nuclei. We cite for 
instance Refs.~\cite{Bender03,Burvenich04,Bonneau04,Sadhukhan14,Schunck14,Ryssens19}
which all report an effect of about 2~MeV on the inner barrier of $^{240}$Pu
using a variety of Skyrme parameterizations. A more systematic study 
found effects on the inner barriers of even-even actinide nuclei that are comparable
to the ones we report~\cite{Ling20}. For EDFs of the Gogny type, allowing for 
triaxial deformation results in a lowering of the inner barrier by about 2~MeV 
for $^{240}$Pu~\cite{Girod83} and up to 4~MeV for other actinide nuclei~\cite{Delaroche06}. 
Two systematic studies employing relativistic models reported effects on the
inner barrier up to about 4~MeV~\cite{Abusara10,Lu2012}.

The study of triaxial deformation near the outer barrier in EDF-based models 
is more recent. The relevance of shapes combining octupole and triaxial 
deformation for actinide fission barriers was first pointed out in the context 
of relativistic models in Refs.~\cite{Lu2012,Lu2014}. The authors report energy 
gains due to triaxiality that grow with neutron number in the range of 0.5 to 
1~MeV, i.~e.\ slightly larger than those shown in Fig.~\ref{fig:effect_triax_outer}. 
For Skyrme-type EDFs, triaxial deformation near the outer barrier was reported 
on for $^{240}$Pu in Ref.~\cite{Ryssens19b} and studied more systematically for 
even-even actinides in Ref.~\cite{Ling20}. The authors of the latter employ the 
UNEDF1 parameterization and find energy gains due to triaxial deformation that 
grow with neutron number and are on the order of a few hundred keV, i.~e.\ similar
in size to those we report on here. To the best of our knowledge, the possibility of 
triaxial deformation near the outer barrier has not yet been studied for Gogny-type EDFs. 

EDF-based approaches thus paint a rather consistent picture with respect to the 
role of triaxiality for the barriers of actinide nuclei when constructing the LEP: 
up to several MeV for the inner barrier and up to one MeV for the outer barrier, 
both effects growing with neutron number. Triaxial deformation seems to be less
 important for non-self-consistent approaches: the lowering of the inner barrier
 for actinide nuclei due to triaxial deformation predicted by such models is typically less than one 
MeV~\cite{Pashkevich69,Larsson72,Gotz72,Dutta00,Moller09,Kowal10,Jachim12}. Ref.~\cite{Jachim12}
is the only mic-mac study of the outer barrier that considers triaxial 
deformation that we are aware of: its authors report an effect of 150 keV on the
outer barrier due to triaxiality for $^{248}$Cm, but did not consider any 
other nuclei. The reduced importance of triaxiality in mic-mac models seems 
consistent with the observation that the gains in g.s. binding energy due to 
triaxial deformation in such models also tend to be smaller than in EDF-based 
approaches~\cite{Scamps21}. 

We remind the reader that our study concerns only static aspects of the fission 
of actinides. A more complete calculation that accounts for dynamical aspects of fission would 
substitute a LAP for each LEP we discuss here. The results of such a study would
likely depend on (i) the treatment of the collective inertia and (ii) the 
selection of collective coordinates. For instance, a perturbative treatment of 
the collective inertia favors an axially symmetric LAP near the inner saddle point 
while the LAP obtained with the so-called cranking approximation for 
the inertia tensor is close to the LEP~\cite{Sadhukhan13}. On the other hand, 
even with this approximation, the inclusion of pairing fluctuations again 
restores axial symmetry along the LAP near the inner saddle point~\cite{Sadhukhan14}. 
However, these conclusions are so far limited in scope: they have only 
been established for the case of $^{240}$Pu and the SkM* parameterization, 
and only for the inner saddle point. Furthermore, they are certainly 
sensitive to the details of the treatment of pairing correlations and 
could perhaps change when more complete treatments of the collective 
inertia are used~\cite{Giuliani18}. 
More work is clearly necessary to extend these studies by (i) including
outer saddle points, (ii) studying more nuclei, particularly those where the effect of 
triaxial deformation on the barriers is larger than for $^{240}$Pu; and (iii) 
improving the treatment of the inertia tensor beyond the cranking approximation.

%%%%%%%%%%%%%%%%%%%%%%%%%%%%%%%%%%%%%%%%%%%%%%%%%%%%%%%%%%%%%%%%%%%%%%%%%%%%%%%%
\subsection{Impact of time-reversal symmetry breaking}
\label{sec:fission_TR}

\begin{figure}
\centering
\includegraphics[width=.45\textwidth]{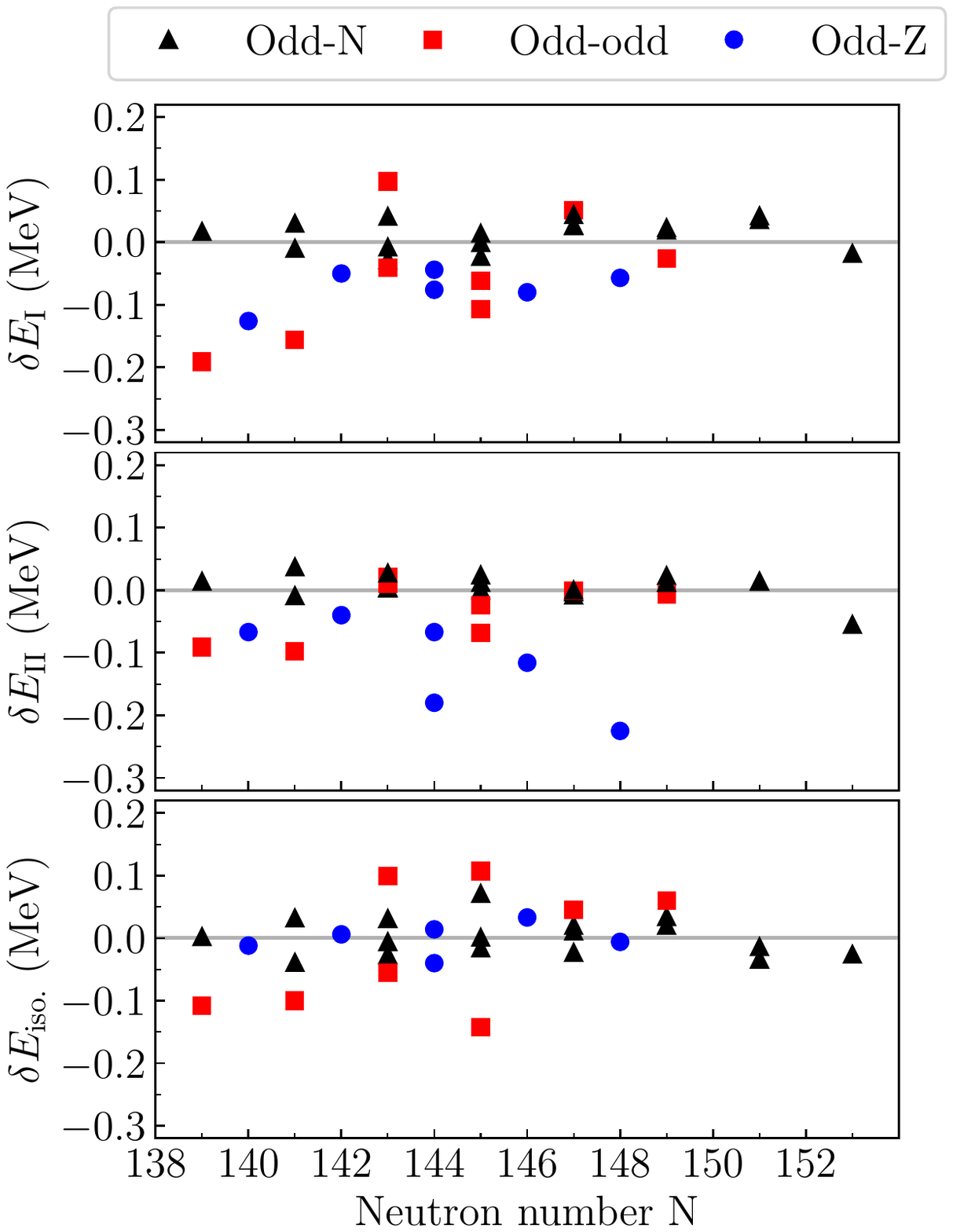}
\caption{The differences of fission properties obtained from an EFA calculation 
          and a full calculation with BSkG2 
          ($\delta E_X = E_X^{\rm Full} - E_{X}^{\rm EFA}$)
          for odd-mass and odd-odd nuclei.
          From top to bottom: primary barriers, secondary barriers and 
          isomer excitation energies. 
        }
\label{fig:timeodd_barriers}
\end{figure}

In order to study the impact of time-reversal symmetry breaking, we also 
performed a full set of fission calculations for BSkG2 with the EFA~\cite{Perez08}. 
As explained in Ref.~\cite{Ryssens22}, treating odd-mass or odd-odd nuclei this way 
allows for simplified calculations yet includes the blocking effect of the 
odd particle(s), neglecting only the influence of the time-odd terms in the EDF. 
Fig.~\ref{fig:timeodd_barriers} shows the difference between a full calculation
that accounts for the time-odd terms and one employing the EFA for all three 
fission properties of the 31 $Z\geq 90$ odd-mass and odd-odd nuclei in our sample.
In all panels, positive (negative) values of $\delta E_{X}$ indicates that
the full calculation produces larger (smaller) values of $E_{\rm X}$ than an 
EFA calculation.

The barriers of odd-$N$, even-$Z$ nuclei are almost entirely 
unaffected by the presence of time-odd terms, but the barriers of most 
systems with an odd number of protons get somewhat lowered. The situation is 
less clear for the isomer excitation energies, which can get lowered or 
enhanced by the time-odd terms. On the whole however, the effect of the 
time-odd terms is small: the largest effect in our set of nuclei is the
lowering of the secondary barrier of $^{243}$Am by about 220 keV while the 
typical effect is below 100 keV for almost all other nuclei. Although this 
could not have been predicted with certainty from our previous study~\cite{Ryssens22}, 
the overall small effect of time-odd 
terms on fission is natural in light of their limited effect on the binding 
energies of heavy nuclei. These results compare rather well with the more 
limited study of Ref.~\cite{Koh17} for $^{239}$Pu and $^{241}$Pu: the authors 
also find a lowering of the inner barrier by a few hundred keV and a less 
systematic effect on the isomer excitation energies. These energy differences
play almost no role in the deviations w.r.t. the empirical values reported
in Table~\ref{tab:barrier_properties}: when calculated with the EFA, the
rms deviations on the primary barriers are 0.42 MeV, 0.51 MeV and 0.27 MeV 
for odd-$Z$, odd-$N$ and odd-odd nuclei respectively.

The chief conclusion of this section is that the impact of time-odd terms on the
fission properties of odd-mass and odd-odd nuclei is on the order of 100~keV. 
This is a small effect when compared to the typical accuracy of BSkG2 with
respect to the empirical values and is particularly negligible when 
considering the conceptual problems surrounding the determination of barriers for
odd-mass and odd-odd nuclei we raised in Sec.~\ref{sec:fission}. In addition, 
for many nuclei the size of the effect is comparable to the numerical accuracy 
of our calculations, see Sec.~\ref{sec:numerical}. For these reasons, we refrain
from any deeper investigation. However, the limited effect of time-odd terms 
on \emph{static} properties does not mean that time-reversal symmetry breaking is 
not relevant to nuclear fission: time-odd terms strongly affect the \emph{dynamic} 
fission properties of all nuclei, including even-even ones. They appear 
naturally in the calculations of the collective inertia~\cite{Matsu10,Baran11,Washiyama21}, 
although their contribution is omitted in the majority of studies published so far.

%%%%%%%%%%%%%%%%%%%%%%%%%%%%%%%%%%%%%%%%%%%%%%%%%%%%%%%%%%%%%%%%%%%%%%%%%%%%%%%%
\subsection{Comparison to other models}
\label{sec:comparison}

\begin{table*}
\centering
%\begin{ruledtabular}
\begin{tabular}{lcccccccccccccccccc}
\hline\noalign{\smallskip}
       &   & \multicolumn{2}{c}{Triaxial}   & &  & \multicolumn{2}{c}{$E_{\rm I}$}   & & \multicolumn{2}{c}{$E_{\rm II}$}   &
                                               &
\multicolumn{2}{c}{$E_{\rm iso}$} & & \multicolumn{2}{c}{$(E_{\rm I} - E_{\rm II})$} \\
\noalign{\smallskip}
\cline{3-4}\cline{7-8}\cline{10-11}\cline{13-14}\cline{16-17}
\noalign{\smallskip}
  Model & Fit &  I & O                 & $N_{\rm b}$& $N_{\rm iso}$  & \multicolumn{1}{c}{$\sigma$} & \multicolumn{1}{c}{$\bar{\epsilon}$} & &
                                                 \multicolumn{1}{c}{$\sigma$} & \multicolumn{1}{c}{$\bar{\epsilon}$} & &
                                                 \multicolumn{1}{c}{$\sigma$} & \multicolumn{1}{c}{$\bar{\epsilon}$} & &
                                                 \multicolumn{1}{c}{$\sigma$} & \multicolumn{1}{c}{$\bar{\epsilon}$} & 
                                                 Ref.\\
\noalign{\smallskip}\hline\noalign{\smallskip}
BSkG1 & N & Y & Y & 45 & 28 & 0.88   & $+$0.80  &&  0.87  &  $+$0.71 &&     1.00  &   $+$0.67  && 0.56 & $+$0.09 & \\
BSkG2 & Y & Y & Y & 45 & 28 & 0.44   & $+$0.24  &&  0.47  &  $+$0.10 &&     0.49  &   $-$0.36  && 0.53 & $+$0.14 & \\
%\hline
% \noalign{\smallskip}
BSk14    & Y & N & N & 45 & 28 & 0.60   & $-$0.27  &&  0.69  &  $+$0.20 &&     1.05  &   $+$0.34  && 0.76  & $-$0.47 &\cite{Goriely07} \\
BCPM     & N & N & N & 45 & 28 & 1.42   & $-$1.07  &&  0.72  &  $-$0.30 &&     0.52  &   $+$0.09  && 1.22  & $-$0.77 &\cite{Giuliani18}\\
SkSC4    & N & N & N & 45 &  0 & 0.57   & $+$0.04  &&  2.03  &  $+$1.78 &&     $-$   &   $-$      && 2.15  & $-$1.74 &\cite{Mamdouh01} \\
FRLDM    & Y & Y & N & 45 & 28 & 0.81   & $+$0.22  &&  1.41  &  $+$0.66 &&     1.02  &   $-$0.91  && 0.88  & $-$0.44 &\cite{Moller09}  \\
YPE+WS   & Y & Y & N & 45 & 28 & 0.82   & $-$0.66  &&  0.84  &  $-$0.40 &&     0.38  &   $+$0.07  && 0.72  & $-$0.26 &\cite{Jachim20}  \\
\noalign{\smallskip}\hline\noalign{\smallskip}
D1M   & Y & Y & N & 14 & 8 & 0.53    & $+$0.23 && 0.43    &  $+$0.06 &&      0.99  & $+$0.50    && 0.47 & +0.17   &\cite{Lemaitre18} \\
UNEDF1& Y & Y & N & 10 & 4 & 0.72    & $-$0.67 && 0.79    &  $-$0.41 &&      0.16  & $-$0.06    && 0.83 & $-$0.26 &\cite{Kortelainen12}\\
      & Y & Y & Y & 12 & 8 & 0.71    & $-$0.52 && 0.65    &  $-$0.28 &&      0.69  & $-$0.36    && 0.71 & $-$0.24 &\cite{Ling20}\\
SkM*  & Y & Y & N & 10 & 0 & 1.92    & $-$1.86 && 1.93    &  $-$1.84 &&      $-$   & $-$        && 0.57 & $-$0.01 &\cite{Kortelainen12}\\
SkI3  & N & N & N &  7 & 8 & 3.99    & $-$3.59 && 1.59    &  $-$1.44 &&      1.04  & $+$0.35    && 2.51 & $-$2.15 &\cite{Burvenich04}\\
      & N & N & N & 14 & 0 & 3.26    & $-$2.50 && $-$     &   $-$    &&       $-$  &  $-$       &&  $-$ &  $-$ & \cite{Erler12} \\ 
SkI4  & N & N & N &  7 & 8 & 4.35    & $-$4.27 && 3.65    &  $-$3.49 &&      0.95  & $-$0.22    && 1.02 & $-$0.78 &\cite{Burvenich04}\\
SLy6  & N & N & N &  7 & 8 & 4.23    & $-$3.90 && 2.19    &  $-$2.08 &&      1.24  & $-$1.28    && 2.24 & $-$1.82 &\cite{Burvenich04}\\ 
      & N & N & N & 14 & 0 & 3.89    & $-$3.31 && $-$     &   $-$    &&       $-$  &  $-$       &&  $-$ &  $-$ & \cite{Erler12} \\ 
SV-bas & N & N & N & 14 &0 & 1.88    & $-$1.10 && $-$     &   $-$    &&       $-$  &  $-$       &&  $-$ &  $-$ & \cite{Erler12} \\ 
SV-min & N & N & N & 14 & 0& 1.61    & $-$0.50 && $-$     &   $-$    &&       $-$  &  $-$       &&  $-$ &  $-$ & \cite{Erler12} \\  
\noalign{\smallskip}\hline\noalign{\smallskip}
NL-Z2  & N & N & N &  7 & 8 & 1.73    & $-$0.93 && 1.28    &  $+$1.19 &&      1.81  & $+$1.91    && 2.68    & $-$2.12 &\cite{Burvenich04}\\
NL3   & N & N & N &  7 & 8 & 2.18    & $-$1.26 && 1.03    &  $+$0.62 &&      0.49  & $+$0.39    && 2.73 & $-$1.88&\cite{Burvenich04}\\
NL3*  & N & N & N & 14 & 0 & 2.16    & $-$2.03 && $-$     &  $-$     &&      $-$   & $-$        && $-$     & $-$&\cite{Taninah20}\\
PC-PK1& N & N & N & 14 & 0 & 1.84    & $-$1.53 && 1.01    &  $-$0.60 &&      $-$   & $-$        && 1.43    & $-$0.93 &\cite{Lu2014} \\ 
      & N & Y & Y & 14 & 0 & 0.37    & $+$0.18 && 0.82    &  $+$0.13 &&      $-$   & $-$        && 0.73    & +0.05   &\cite{Lu2014}\\ 
DD-ME2& N & N & N & 14 & 0 & 3.35    & $-$3.17 && $-$     &  $-$     &&      $-$   & $-$        && $-$& $-$&\cite{Taninah20}\\
DD-PC1& N & N & N & 14 & 0 & 2.45    & $-$1.76 && $-$     &  $-$     &&      $-$   & $-$        && $-$& $-$&\cite{Taninah20}\\
\noalign{\smallskip}\hline
\end{tabular}
%\end{ruledtabular}
\caption{  Rms deviations ($\sigma$) and mean ($\bar{\epsilon}$) errors
           ($\delta E_{\rm X} = E_{\rm X}^{\text{emp}} - E_{\rm X}^{\rm calc}$)
           for the height of the primary $(E_{\rm I})$ and secondary $(E_{\rm II})$ barriers, 
           the excitation energy of the isomer ($E_{\rm iso}$) and the
           barrier difference $(E_{\rm I} - E_{\rm II})$
           for various models. The second column indicate whether the 
           models parameter adjustment included information on the physics
           of large deformation, in one form or another.
           The third and fourth column indicate whether
           or not the calculations considered triaxial deformation near
           the inner and outer barrier respectively. $N_{\rm b}$ and $N_{\rm iso}$ 
           respectively refer to the number of barriers and isomers included in the 
           calculated deviations.  
          }
\label{tab:comparison_model}
\end{table*}

We have compiled in Table~\ref{tab:comparison_model} the rms and mean deviations
with respect to the empirical values for the barriers and isomer excitation 
energies for different models available in the literature. The first group 
contains all models that cover the full extent of all 45 nuclei with $Z \geq 90$
in the RIPL-3 database. This group contains four EDF-based models: BSkG1, BSkG2,
BSk14~\cite{Goriely07} and BCPM~\cite{Giuliani18}. It also includes three 
non-self-consistent models: the ETFSI calculations based on the SkSC4 
interaction~\cite{Mamdouh01} and the mic-mac FRLDM~\cite{Moller09}
and YPE+WS~\cite{Jachim20,Jachimowicz21} models. FRLDM and YPE+WS both employ
a very similar Yukawa-plus-exponential (YPE) macroscopic part, but combine it with
different microscopic ingredients: a Folded-Yukawa model for the former and a 
Woods-Saxon (WS) model for the latter.

The second and third groups consist of models whose predictions cover at least 
7 even-even nuclei out of 14 in the original selection. Compared to global models, there are much more
works discussing the fission properties of a limited amount of even-even 
actinide nuclei, such that we have limited ourselves to EDF-based models. The second
group consists of non-relativistic models: Gogny-type D1M~\cite{Lemaitre18} 
and Skyrme-type UNEDF1~\cite{Kortelainen12,Ling20}, SkI3~\cite{Burvenich04}, 
SkI4~\cite{Burvenich04}, SLy6~\cite{Burvenich04} and SkM*~\cite{Kortelainen12}
\footnote{Other sets of fission data obtained with SkM* exist, such as  
   Ref.~\cite{Bonneau04}, but Ref.~\cite{Kortelainen12} reports the most 
   extensive data set that we are aware of.}.
Although the calculations in Ref.~\cite{Erler12} were extensive, they concern
only even-even nuclei such that we are forced to include their Skyrme results for
the SkI3, SLy6, SV-bas and SV-min parameterizations in this second group.  
The third group consists of relativistic models: NL-Z2~\cite{Burvenich04}, 
NL3~\cite{Burvenich04}, PC-PK1~\cite{Lu2014} and the set DD-ME2, DD-PC1 and 
NL3* of Ref.~\cite{Taninah20}. NL3 and NL3* are closely related, the latter
is a more modern refit of the original~\cite{Lalazissis97,Lalazissis09}.
For some data sets, isomer excitation energies or secondary barrier heights were not available. 

In four cases, we list two different sets of predictions: for UNEDF1 we show the
values obtained in Ref.~\cite{Ling20} which considered triaxial deformation
near both saddle points and the values of the original paper~\cite{Kortelainen12}
which considered triaxiality only near the first saddle point. 
For PC-PK1, we include the complete results of Ref.~\cite{Lu2014}: the calculations 
limited to axial symmetry and those considering triaxial deformation. 
For SkI3 and SLy6 we include the results of both Ref.~\cite{Burvenich04} 
and the more recent Ref.~\cite{Erler12}: the results of the latter cover more
nuclei, but for the former data on secondary barriers and isomer excitation 
energies were also available.

Table~\ref{tab:comparison_model} also notes which models allowed for triaxial 
deformation: BSkG1 and BSkG2 are the only EDF-based models in the first group 
that account for it near both saddle points, while the mic-mac FRLDM and 
YPE+WS models include its effect only near the inner barrier. Several models in 
the second and third groups considered triaxial deformation near the inner barrier, 
but only two (PC-PK1 and UNEDF1) incorporated it near the second barrier. 
All models in Table 3 account for octupole deformation near the outer saddle 
point since its effect on the outer barrier is so large for 
actinide nuclei. For further details of the strong points and flaws of each 
model, we refer the readers to the original references.

The second column of Table~\ref{tab:comparison_model} shows whether or not
the parameter adjustment of the model included, in one form or another, the
physics of large deformation. For the Skyrme models  BSkG2, BSk14, UNEDF1 and 
SkM*, we already discussed how this was done in Sec.~\ref{sect:parameter:adjustment}. 
The surface properties of the D1M Gogny model were adjusted along the same
lines as those of the earlier D1S~\cite{Goriely09b}. The objective function 
of the parameters of the macroscopic energy of the FRLDM model included 31 
fission barriers~\cite{Moller16}. The YPE+WS model employs the macroscopic
part of an earlier version of the FRLDM model~\cite{Muntian01}, 
which was also adjusted to 28 barriers~\cite{Moller95}. 
Although no data on fission properties directly entered the fit of any of
the selected relativistic models, we note that the surface properties
of DD-PC1 were carefully constrained on the masses of deformed heavy nuclei~\cite{Niksic08}.

Table~\ref{tab:comparison_model} regroups a large diversity of models with 
a corresponding diversity in numerical accuracy. For the numerical conditions
outlined in Sec.~\ref{sec:numerical}, we estimate that our calculations 
reach a numerical accuracy of about 100 keV that is independent of the
nuclear shape and thus applies to the inner saddle point, isomer and outer
saddle point equally. The results for NL3, NL-Z2, SkI3, SkI4 and SLy6 of 
Ref.~\cite{Burvenich04} and the SkI3, SLy6, SV-bas and SV-min results of Ref.~\cite{Erler12} 
were obtained with a coordinate-space approach similar to ours: in the case of the 
former their choice of mesh spacing ($1$ fm) leads to an accuracy that 
is slightly lower than ours~\cite{Ryssens15b}. The accuracy of both sets of 
results nevertheless remains independent of elongation~\cite{Ryssens15b}. 
Although we are not aware of any dedicated studies, the numerical 
accuracy of mic-mac approaches is likely even better as most ingredients of
the macroscopic part of the energy are analytical functions of the shape
parameters\footnote{Only the Coulomb energy and surface energy require
the numerical evaluation of an integral. Nevertheless, the numerical accuracy
of these terms should be easily controllable~\cite{Moller16}.}.
However, all other results in Table~\ref{tab:comparison_model} were obtained
using numerical implementations relying on an expansion in a limited number
of harmonic oscillator states. The numerical accuracy of such implementations
is typically dependent on deformation since the harmonic oscillator spectrum
is not well suited to represent very elongated shapes. Ref.~\cite{Lu2014}, 
for example, cites a numerical accuracy of about 150 keV and 400 keV for the 
inner and outer barriers obtained with PC-PK1. Ref.~\cite{Karatzikos10} cites
a similar 100 keV accuracy on the inner barriers for the conditions used 
in Ref.~\cite{Taninah20}. In light of this issue, we will restrict 
ourselves to more qualitative observations of the whole of Table~\ref{tab:comparison_model}.  

Correlating the third columns of Table~\ref{tab:comparison_model} with the 
seventh and eight columns shows that models that include triaxial deformation
near the inner saddle barrier systematically outperform models that do not. 
It is also clear, but unsurprising, that models that incorporate the physics of 
large deformation in their parameter adjustment typically do better than 
others constructed without such information: the older SkM* parameterization
is the only model with an informed fit that does not reach 1 MeV accuracy on 
the primary barriers even when allowing for triaxial deformation near the 
inner saddle point. The relativistic PC-PK1 parameterization is an 
interesting outlier: its rms deviation for the primary barriers of 14 even-even 
nuclei is the lowest of all models in Table~\ref{tab:comparison_model}, yet 
its parameter adjustment considered only spherical nuclei~\cite{Zhao10}. 
If triaxial deformation is not accounted for \emph{and}  
large-deformation physics is not present during parameter adjustment, the 
resulting self-consisten models systematically overestimate the primary barriers.

BSkG2 can be compared directly with BSk14: the latter also includes both 
a rotational and vibrational correction whose parameters were adjusted to 
fission data in a two-step procedure similar to ours. For the primary barriers, 
arguably the most important among the three fission properties, BSkG2 improves 
on the BSk14 rms deviation by about 20 percent. BSkG2 also offers rms deviations
on the secondary barriers and isomer excitation energies that are, respectively, 
more than 30 and 50 percent lower than those of BSk14. Also worthy of discussion
are the mean deviations: the BSkG2 mean deviations for $E_{\rm I}, E_{\rm II}$
and $E_{\rm iso}$ are comparable in size to those of BSk14, but the difference
$\delta E_{\rm I, II}$ is much more accurately described by BSkG2, as 
discussed above.

Table~\ref{tab:comparison_model} also shows that \emph{all} models that do not
include triaxial deformation systematically overestimate $E_{\rm I} - E_{\rm II}$.
Including triaxial deformation near the first barrier is most important; 
self-consistent models that do this all achieve 
$\left|\epsilon(E_{\rm I} - E_{\rm II})\right| \leq 0.26 $ MeV 
while ones that do not have mean deviations that are at least twice as large. 
Even BSkG1 achieves an accurate description of $E_{\rm I} - E_{\rm II}$,
despite its systematic underestimation of both barriers. This indicates again
that the difference $(E_{\rm I} - E_{\rm II})$ is sensitive to triaxial deformation and 
does not depend significantly on the treatment of the collective correction. 

The most important quantity for applications remains the primary barrier. 
In that category, BSkG2 has the lowest rms deviation among all models in the 
first group. Among the second group, only D1M and PC-PK1 (with triaxial 
deformation) offer similar accuracy. Several models achieve a mean deviation 
for $E_{\rm I}$ that is comparable to that of BSkG2, but only SkSC4 does 
significantly better. We could have further improved slightly the performance of BSkG2 for the 
barriers by fine-tuning the collective correction more, but only at the 
cost of our description of the isomer excitation energies. However, the philosophy of the BSkG
models and the older BSk models is to describe \emph{simultaneously} 
as many (pseudo-)observables as possible in order to maximise predictive
power. If we also take into account the secondary barriers and fission isomers, 
it is clear that BSkG2 is the most accurate `all-round' fission model. While
there are several models that can compete with BSkG2 in any given column of 
Tab.~\ref{tab:comparison_model}, most of them perform significantly worse
in one or more other categories. The most extreme example is SkSC4, 
which offers one of the smallest mean deviations on the primary barriers among all
models studied here but also massively underestimates the secondary barriers.
D1M essentially matches the performance of BSkG2 for barriers, but not for isomers. 
PC-PK1 does somewhat better for primary barriers, but somewhat worse for secondary
barriers. We remind the reader that for both D1M and PC-PK1, the
deviations are evaluated with respect to a much smaller set of nuclei.

Finally, a remark on the number of parameters is in order: the BSkG2 model is
formulated in terms of 25 parameters, somewhat more than the 22 of BSkG1 and 
the 24 of BSk14. This is significantly more parameters than most other EDF-based
models in Tab.~\ref{tab:comparison_model} have; for instance PC-PK1 and D1M
depend on 11 and 14 parameters, respectively. Although a smaller amount of
parameters constitutes an advantage for any model, one should keep in mind that 
we focussed here solely on fission properties: (i) BSkG2 describes several other
properties of nuclei more accurately than many of the other EDF-based models in 
Tab.~\ref{tab:comparison_model} and (ii) a significant number of model parameters impact the fission 
properties only tangentially (e.g. the spin-orbit strengths $W_0$ and $W'_0$) 
or even not at all (e.g. the four parameters of the Wigner energy).

%%%%%%%%%%%%%%%%%%%%%%%%%%%%%%%%%%%%%%%%%%%%%%%%%%%%%%%%%%%%%%%%%%%%%%%%%%%%%%%%
\section{Conclusion and outlook}
\label{sec:conclusions}

\subsection{Conclusions}

We have presented the fission properties of the recent BSkG2 model, focussing 
on the effect of triaxial deformation and the effect of 
time-odd terms on the barrier heights and isomer excitation energies of the 
double-humped fission barrier of actinide nuclei. We presented results for all 
$Z \geq 90$ nuclei for which empirical barriers are available in the 
RIPL-3 database~\cite{Capote09} and described in detail our calculations: 
they are three-dimensional and simultaneously allow for octupole deformation, 
time-reversal symmetry breaking and triaxial deformation. The use of a 
coordinate-space representation of the single-particle wave functions results in
a high numerical accuracy, typically about 100 keV.

We benchmarked the lowering of the inner and outer barrier due to triaxial
deformation: all 45 nuclei in our sample are affected by up to 4 MeV 
for the inner barrier and up to 0.8 MeV for the outer barrier. The effect grows 
quickly with increasing neutron number for both inner and outer barriers, 
suggesting the importance of neutron shell effects and motivating the extension 
of our results to more neutron-rich nuclei. Although the effect on the barriers
is large, the fission path in terms of the nuclear shape is not much 
affected. Time-odd terms on the other hand do not influence much the 
static fission properties we discuss here; their effect on the barriers
and isomer excitation energies can be of either sign, but its size is typically
less than 100 keV. At the time of writing, this contribution is the largest study
in the context of fission of triaxial deformation and time-odd terms separately, 
and is the only study to combine both. 

The main conclusion is that BSkG2 achieves an excellent description of the 
primary and secondary barriers as well as the isomer excitation energies of 
actinide nuclei, reaching rms deviations for all three quantities below 500 keV. 
Although partially due to the inclusion of fission properties
in the parameter adjustment, the inclusion of triaxial deformation was crucial
to achieve this level of accuracy. In particular, the new model successfully 
describes the difference between the primary and secondary barriers in the actinide 
region; models that do not incorporate triaxial deformation systematically 
overestimate this quantity.

We have compared the performance of BSkG2 to a large selection of models in 
the literature: BSkG2 offers the best description of primary barriers
among all large-scale fission models for which values for all actinide nuclei
are available. With the exception of PC-PK1~\cite{Lu2014} and D1M~\cite{Lemaitre18}, 
BSkG2 also outperforms most models for which only predictions for 
even-even nuclei are available. However, the \emph{combined} accuracy of 
BSkG2 on the primary and secondary barriers as well as the isomer excitation 
energies is unrivalled in the published literature. 

In the preceding paper, Ref.~\cite{Ryssens22}, we established that BSkG2
offers both a state-of-the-art description of g.s. properties and reasonable
predictions for infinite nuclear matter. These qualities have now been combined
with an unprecedented performance with respect to actinide fission barriers,
rendering the model uniquely suited to provide data to nuclear applications in 
general and r-process calculations in particular.

\subsection{ Outlook }
While the present study already extensively covers the actinides with known 
barriers, the description of a few tens of nuclei is not our ultimate aim: what is 
required for r-process simulations are the fission rates and fragment yields 
of thousands of neutron-rich nuclei for many different fission channels. 
This study is the first step on the way there: without a proper reproduction of 
experimental data and a detailed understanding of the successes and 
flaws of the BSkG models, we cannot have confidence in their predictions for 
exotic nuclei. Getting the BSkG-series to the desired point will require at least further technical 
developments for the calculation of inertial masses~\cite{Baran11,Giuliani18}, 
saddle-point level densities~\cite{Goriely08b} and fragment 
yields~\cite{Schmidt16, Lemaitre19,Lemaitre21}, as well as an extension of all 
these efforts to thousands of heavy (and mostly neutron-rich) nuclei. As discussed in 
Sec.~\ref{sec:triax} and Sec.~\ref{sec:fission_TR}, the details of our
treatment of the collective inertia might change our conclusions on the role of 
triaxial deformation and time-reversal symmetry breaking.

Aside from these extensions, we see two main weak points in our description
of nuclear fission. The first is conceptual and concerns the determination of the
barrier of odd-mass and odd-odd nuclei: can the traditional semi-classical view of fission as large-scale motion 
in a limited set of collective variables be applied to such systems? More precisely, 
it is not clear how to construct a fission path for such nuclei that connects 
each blocked HFB configuration to the next in a continuous way. 
Nor is it clear how one should choose the relevant blocked HFB configuration. 
For example, how is the angular momentum in the g.s. to be treated: via the 
construction of several different fission paths for each value of the $K$ quantum number
as in Refs.~\cite{Koh17,Schunck22x}, even when it loses its meaning
when allowing for triaxial deformation? Furthermore, is an adiabiatic evolution 
along the PES always appropriate? Given the high level density at low energy for odd-mass
and odd-odd nuclei, should one incorporate the possibility of diabatic transitions
for the blocked quasiparticles? The latter is also an open, though perhaps
less pressing, question for even-even systems~\cite{Bender20}.

The second weak point is imposed on us by computational considerations:
our approach to collective motion and in particular the vibrational
correction remains highly phenomenological out of necessity. As we remarked already in 
Refs.~\cite{Scamps21} and \cite{Ryssens22}, symmetry-restoration techniques 
should be adopted to correct the deficiencies of our symmetry-broken approach 
which has no access to quantum numbers. Applying such techniques to 
the ground states of all nuclei is a significant challenge, but incorporating 
them into a description of fission is even more difficult. Although some 
works of more limited scope exist~\cite{Samyn05,Hao12,Bernard19}, rotational 
symmetry restoration has only been applied to the fission barrier of a single 
even-even nucleus: $^{240}$Pu~\cite{Bender98,Marevic20}. Pending significant
further developments in this direction, microscopically motivated approximations
could be devised based on, for example, replacing the Belyaev MOI in 
Eq.~\eqref{eq:Erot} by the rotational and vibrational inertia obtained from
a linear-response calculation. Employing linear-response techniques would 
also naturally provide access to a microscopic treatment of the collective inertia
that is required for the determination of the LAP~\cite{Dobaczewski19,Bender20}.

\begin{acknowledgements}
We would like to thank A. Afanasjev, S. Giuliani and P.-G. Reinhard for providing us 
with additional data on their calculations for use in Sec.~\ref{sec:comparison}.: 
these concern Ref.~\cite{Taninah20}, ~\cite{Giuliani18} and \cite{Erler12}, respectively.
We also thank J.-F. Lemaître for sharing his flooding code.
W.R.\ acknowledges useful discussions with N.~Schunck.
This work was supported by the Fonds de la Recherche Scientifique (F.R.S.-FNRS) 
and the Fonds Wetenschappelijk Onderzoek-Vlaanderen (FWO) under the EOS 
Project nr O022818F. The present research benefited from computational resources 
made available on the Tier-1 supercomputer of the F\'ed\'eration 
Wallonie-Bruxelles, infrastructure funded by the Walloon Region under the grant 
agreement nr 1117545. 
The funding for G.S.\ from the US DOE, Office of Science, Grant No. DE-FG02-97ER41014 is greatly appreciated.
S.G.\ and W.R.\ acknowledge financial support from the 
F.R.S.-FNRS (Belgium). Work by M.B.\ has been supported by the French Agence Nationale 
de la Recherche under grant No.\ 19-CE31-0015-01 (NEWFUN).
\end{acknowledgements}

\section{Explanation of the supplementary material}

We provide \textsf{Fission\_Table\_BSkG2.dat} as supplementary material.
It contains the fission barriers and isomer excitation energies as calculated
with BSkG2 for all 45 actinide nuclei considered in the text, including their location 
on the PES in terms of quadrupole ($\beta_{20}, \beta_{22}$) and octupole
deformation ($\beta_{30}$). For convenience, Table~\ref{tab:suppl} contains
an explanation of all the columns of the file, grouped by fission property. 
For all nuclei, the isomer is axially and reflection symmetric 
($\beta_{22} = \beta_{30} = 0$) while the inner barrier is reflection symmetric 
($\beta_{30} = 0$); the corresponding multipole moments have not been 
included in the table.

\begin{table*}[]
\centering
\begin{tabular}{lllll}
\hline
\hline
Column &  Quantity & Fission property & Units & Explanation   \\
\hline
1 & Z & $-$ &$-$ & Proton number  \\
2 & N & $-$ &$-$ & Neutron number \\
\hline
%  & Inner barrier & & \\
3 & $E$          & Inner barrier  & MeV  & Barrier height          \\
4 & $\beta_{20}$ &                & $-$  & Quadrupole deformation  \\
5 & $\beta_{22}$ &                & $-$  &                         \\
\hline
%  & Outer barrier & & \\
6 & $E$          & Outer barrier  & MeV & Barrier height           \\
7 & $\beta_{20}$ &                & $-$ & Quadrupole deformation   \\
8 & $\beta_{22}$ &                & $-$ & \\
9 & $\beta_{30}$ &                & $-$ & Octupole deformation \\
\hline
10 & $E$         & Isomer         & MeV &  Excitation energy     \\
11 & $\beta_{20}$&                & $-$ &  Quadrupole deformation    \\
\hline 
\hline
\end{tabular}
\caption{Contents of the \textsf{Fission\_Table\_BSkG2.dat} file.}
\label{tab:suppl}
\end{table*}


\begin{thebibliography}{100}

\bibitem{Arnould20} M. Arnould and S. Goriely, 
\emph{Astronuclear Physics: A tale of the atomic nuclei in the skies}, 
Prog. Part. Nucl. Phys. \textbf{112}, 103766 (2020). 
%https://doi.org/10.1016/j.ppnp.2020.103766

%-------------------------------------------------------------------------------
% Mass models

\bibitem{Weizsacker35} C. F. v. Weizs{\"a}cker, 
\emph{Zur Theorie der Kernmassen},
Z. Phys.  \textbf{96}, 431-458 (1935).
% https://doi.org/10.1007/BF01337700

\bibitem{Duflo95} J. Duflo and A. P. Zuker, 
\emph{Microscopic mass formulas},
Phys. Rev. C \textbf{52}, 23-27 (1995). 
%https://doi.org/10.1103/PhysRevC.52.R23

\bibitem{Moller88} P. M\"oller and J.R. Nix, 
\emph{Nuclear masses from a unified macroscopic-microscopic model}, 
At. Data Nucl. Data Tables {\bf 39}, 213 (1988).

\bibitem{Moller95} P. M\"oller, J. R. Nix, W. D. Myers, and W. J. Swiatecki, 
\emph{Nuclear Ground-State Masses and Deformations}, 
At. Data Nucl. Data Tables \textbf{59}, 185 (1995).

\bibitem{Moller16} P. M\"oller, A. J. Sierk, T. Ichikawa and H. Sagawa, 
\emph{Nuclear ground-state masses and deformations: FRDM(2012)}, 
At. Data Nucl. Data Tables, \textbf{109}, 1-204. 
%https://doi.org/10.1016/j.adt.2015.10.002

\bibitem{Tondeur78} F. Tondeur, 
\emph{An Energy Density Nuclear Mass Formula (I). Self-consistent calculation for spherical nuclei},
Nucl. Phys. A \textbf{303}, 185-198 (1978).

\bibitem{Goriely09b} S. Goriely, S. Hilaire, M. Girod and S. P{\'e}ru, 
\emph{First Gogny-Hartree-Fock-Bogoliubov Nuclear Mass Model}, 
Phys. Rev. Lett. \textbf{102}, 242501 (2009).
% https://doi.org/10.1103/PhysRevLett.102.242501

\bibitem{Goriely16} S. Goriely, N. Chamel and J. M. Pearson, 
\emph{Further explorations of Skyrme-Hartree-Fock-Bogoliubov mass formulas. 
         XVI. Inclusion of self-energy effects in pairing},
Phys. Rev. C \textbf{93}, 034337 (2016).

%-------------------------------------------------------------------------------

\bibitem{Pena16} D. Pe\~na-Arteaga, S. Goriely, and N. Chamel, 
\emph{Relativistic mean-field mass models},
Eur. Phys. J. A \textbf{52}, 320 (2016).

\bibitem{Scamps21} G. Scamps, S. Goriely, E. Olsen, M. Bender, and W. Ryssens, 
\emph{Skyrme-Hartree-Fock-Bogoliubov Mass Models on a 3D Mesh: Effect of Triaxial Shape},
Eur. Phys. J. A \textbf{57}, 333 (2021).

\bibitem{Ryssens22} W. Ryssens, G. Scamps, S. Goriely and M. Bender,
\emph{Skyrme-Hartree-Fock-Bogoliubov mass models on a 3D mesh: II. time-reversal symmetry breaking},
Eur. Phys. J. A \textbf{58}, 246 (2022).

\bibitem{Yuksel21} E. Yüksel, D. Soydaner and H. Bahtiyar, 
\emph{Nuclear binding energy predictions using neural networks: Application of the multilayer perceptron},
Int. J. Mod. Phys. E \textbf{30}, 2150017 (2021).
% https://doi.org/10.1142/S0218301321500178

\bibitem{Gazula92} S. Gazula, J. W. Clark and H. Bohr, 
\emph{Learning and prediction of nuclear stability by neural networks}, 
Nucl. Phys. A \textbf{540}, 1-26 (1992).
% https://doi.org/10.1016/0375-9474(92)90191-L

\bibitem{Wang11} N. Wang and M. Liu, 
\emph{Nuclear mass predictions with a radial basis function approach},
Phys. Rev. C \textbf{84}, 051303 (2011).
% https://doi.org/10.1103/PhysRevC.84.051303

\bibitem{Wang14} N. Wang, M. Liu, X. Wu and J. Meng, 
\emph{Surface diffuseness correction in global mass formula}, 
Phys. Lett. B \textbf{734}, 215–219 (2014).
% https://doi.org/10.1016/j.physletb.2014.05.049

\bibitem{Utama17} R. Utama and J. Piekarewicz, 
\emph{Refining mass formulas for astrophysical applications: A Bayesian neural network approach},
Phys. Rev. C \textbf{96}, 044308 (2017). 
%https://doi.org/10.1103/PhysRevC.96.044308

\bibitem{Neufcourt18} L. Neufcourt, Y. Cao, W. Nazarewicz and F. Viens,
\emph{Bayesian approach to model-based extrapolation of nuclear observables}, 
Phys. Rev. C \textbf{98}, 034318 (2018).
%https://doi.org/10.1103/PhysRevC.98.034318

\bibitem{Shelley21} M. Shelley and A. Pastore,
\emph{A New Mass Model for Nuclear Astrophysics: Crossing 200 keV Accuracy},
Universe, \textbf{7}, 131 (2021).
% https://doi.org/10.3390/universe7050131

\bibitem{Ziu22} Z. M. Niu and H. Z. Liang, 
 \emph{Nuclear Mass Predictions with Machine Learning Reaching the Accuracy Required by $r$-Process Studies}, 
Phys. Rev. C \textbf{106}, L021303 (2022).

\bibitem{Ye22} W. Ye, Y. Qian, and Z. Ren, 
 \emph{Accuracy versus Predictive Power in Nuclear Mass Tabulations}, 
 Phys. Rev. C \textbf{106}, 024318 (2022).

\bibitem{Fowler60} W. A. Fowler and F. Hoyle, 
\emph{Nuclear Cosmochronology}, 
Annals of Physics \textbf{10}, 280 (1960).

\bibitem{Goriely01} S. Goriely and M. Arnould, 
\emph{Actinides: How Well Do We Know Their Stellar Production?}, 
Astron. Astrophys. \textbf {379}, 1113 (2001).

\bibitem{Goriely15c} S. Goriely, 
\emph{The fundamental role of fission during r-process nucleosynthesis in neutron star mergers},
Eur. Phys. J. A \textbf{51}, 22 (2015). 
%https://doi.org/10.1140/epja/i2015-15022-3

\bibitem{Bjornholm80} S. Bjørnholm and J. E. Lynn, 
\emph{The Double-Humped Fission Barrier}, 
Rev. Mod. Phys. \textbf{52}, 725 (1980).

\bibitem{Krappe12a} H. Krappe and K. Pomorski, 
\emph{Theory of Nuclear Fission},
(Springer, 2012).

\bibitem{Schunck16a} N. Schunck and L. M. Robledo,
N. Schunck and L. M. Robledo, 
\emph{Microscopic theory of nuclear fission: A review}, 
Rep. Prog. Phys. \textbf{79}, 116301 (2016).

\bibitem{Schmidt18a} K.-H. Schmidt and B. Jurado,
\emph{Review on the progress in nuclear fission-experimental methods and theoretical descriptions},
Rep. Prog. Phys. \textbf{81}, 106301 (2018).

\bibitem{Schunck22a} N. Schunck and D. Regnier,
\emph{Theory of nuclear fission},
Prog. Part. Nucl. Phys. \textbf{125}, 103963 (2022).

\bibitem{Bender20} M. Bender \emph{et al.},  
\emph{Future of nuclear fission theory}, 
J. Phys. G \textbf{47}, 113002 (2020).
% https://doi.org/10.1088/1361-6471/abab4f

\bibitem{Mamdouh01} A. Mamdouh, J. M. Pearson, M. Rayet and F. Tondeur, 
\emph{Fission barriers of neutron-rich and superheavy nuclei calculated with the ETFSI method}, 
Nucl. Phys. A \textbf{679}, 337-358 (2001).
% https://doi.org/10.1016/S0375-9474(00)00358-4

\bibitem{Giuliani13} S. A. Giuliani and L. M. Robledo, 
\emph{Fission properties of the Barcelona-Catania-Paris-Madrid energy density functional}, 
Phys. Rev. C \textbf{88}, 054325 (2013). 
%https://doi.org/10.1103/PhysRevC.88.054325

\bibitem{Bender03} M. Bender, P.-H. Heenen and P.G. Reinhard, 
\emph{Self-consistent mean-field models for nuclear structure},
Rev. Mod. Phys. \textbf{75}, 121 (2003).

\bibitem{Ryssens16} W. Ryssens, 
\emph{Symmetry breaking in nuclear mean-field models}, 
Ph.D. thesis, Universit\'e Libre de Bruxelles, 2016.

\bibitem{Goriely07} S. Goriely, M. Samyn, and J. M. Pearson, 
\emph{Further explorations of Skyrme-Hartree-Fock-Bogoliubov mass formulas. 
VII. Simultaneous fits to masses and fission barriers},
Phys. Rev. C \textbf{75}, 064312 (2007).

\bibitem{Kortelainen15} M. Kortelainen, N. Hinohara and W. Nazarewicz, 
\emph{Multipole modes in deformed nuclei within the finite amplitude method}, 
Phys. Rev. C \textbf{92}, 051302 (2015).
% https://doi.org/10.1103/PhysRevC.92.051302

\bibitem{Jodon16} R. Jodon, M. Bender, K. Bennaceur and J. Meyer, 
\emph{Constraining the surface properties of effective Skyrme interactions},
Phys. Rev. C \textbf{94}, 024335 (2016).
% https://doi.org/10.1103/PhysRevC.94.024335

\bibitem{Capote09} R. Capote \emph{et al}.,
\emph{RIPL - Reference Input Parameter Library for Calculation of Nuclear Reactions and Nuclear Data Evaluation},
Nucl. Data Sheets \textbf{110}, 3107-3214 (2009).
% https://doi.org/10.1016/j.nds.2009.10.004

\bibitem{Wan21} M. Wang \textit{et al.}, 
\emph{The AME 2020 atomic mass evaluation},
Chin. Phys. C \textbf{45}, 3 (2021).

\bibitem{Ryssens15b} W. Ryssens, P.-H. Heenen and M. Bender, 
\emph{Numerical accuracy of mean-field calculations in coordinate space},
Phys. Rev. C \textbf{92}, 064318 (2015).

\bibitem{Girod83} M. Girod and B. Grammaticos,
\emph{Triaxial Hartree-Fock-Bogolyubov calculations with D1 effective interaction},
Phys. Rev. C \textbf{27}, 2317-2339 (1983). 
% https://doi.org/10.1103/PhysRevC.27.2317

\bibitem{Abusara10} H. Abusara, A. V. Afanasjev and P. Ring, 
\emph{Fission barriers in actinides in covariant density functional theory: The role of triaxiality},
Phys. Rev. C \textbf{82}, 044303 (2010). 
%https://doi.org/10.1103/PhysRevC.82.044303

\bibitem{Lu2014} B. N. Lu, J. Zhao, E. G. Zhao and S. G. Zhou, 
\emph{Multidimensionally-constrained relativistic mean-field models and potential-energy surfaces of actinide nuclei}, 
Phys. Rev. C \textbf{89}, 014323 (2014).
% https://doi.org/10.1103/PhysRevC.89.014323

\bibitem{Ryssens19b} W. Ryssens, M. Bender, K. Bennaceur, P.-H. Heenen and J. Meyer, 
\emph{Impact of the surface energy coefficient on the deformation properties of atomic nuclei as predicted by Skyrme energy density functionals}, 
Phys. Rev. C \textbf{99}, 044315 (2019).
% https://doi.org/10.1103/PhysRevC.99.044315

\bibitem{Ling20} C. Ling, C. Zhou, and Y. Shi, 
\emph{Fission Barriers of Actinide Nuclei with Nuclear Density Functional Theory: Influence of the Triaxial Deformation}, 
Eur. Phys. J. A \textbf{56}, 180 (2020).

\bibitem{Cwiok96} S. Ćwiok, J. Dobaczewski, P.-H. Heenen, P. Magierski, and W. Nazarewicz, 
\emph{Shell Structure of the Superheavy Elements}, 
Nucl. Phys. A \textbf{611}, 211 (1996).

\bibitem{Bender98} M. Bender, K. Rutz, P.-G. Reinhard, J. A. Maruhn and W. Greiner, 
\emph{Potential energy surfaces of superheavy nuclei},
 Phys. Rev. C \textbf{58}, 2126-2132 (1998). 
 %https://doi.org/10.1103/PhysRevC.58.2126

\bibitem{Warda02} M. Warda, J. L. Egido, L. M. Robledo and K. Pomorski,
\emph{Self-consistent calculations of fission barriers in the Fm region}, 
Phys. Rev. C  \textbf{66}, 143101 (2002). 
%https://doi.org/10.1103/PhysRevC.66.014310

\bibitem{Abusara12} H. Abusara, A.V. Afanasjev and P. Ring,
\emph{Fission barriers in covariant density functional theory: Extrapolation to superheavy nuclei},
Phys. Rev. C \textbf{85}, 024314 (2012).
% https://doi.org/10.1103/PhysRevC.85.024314

\bibitem{Tondeur00} F. Tondeur, S. Goriely, J. M. Pearson and M. Onsi, 
\emph{Towards a Hartree-Fock mass formula},
Phys. Rev. C \textbf{62}, 024308 (2000).

\bibitem{Goriely13b} S. Goriely, N. Chamel, and J. M. Pearson, 
\emph{Further explorations of Skyrme-Hartree-Fock-Bogoliubov mass formulas. 
XIII. The 2012 atomic mass evaluation and the symmetry coefficient}, 
Phys. Rev. C \textbf{88}, 024308 (2013).

\bibitem{Bender00} M. Bender, K. Rutz, P.-G. Reinhard and J. A. Maruhn, 
\emph{Consequences of the center-of-mass correction in nuclear mean-field models},
Eur. Phys. J. A \textbf{7}, 467-478 (2000).

\bibitem{Goriely03} S. Goriely, M. Samyn, M. Bender and J. M. Pearson, 
\emph{Further explorations of Skyrme-Hartree-Fock-Bogoliubov mass formulas.
     II. Role of the effective mass},
    Phys. Rev. C  \textbf{68}, 054325 (2003).

\bibitem{Sadhukhan14} J. Sadhukhan, J. Dobaczewski, W. Nazarewicz, J. A. Sheikh and A. Baran, 
\emph{Pairing-induced speedup of nuclear spontaneous fission}, 
Phys. Rev. C \textbf{90}, 061304 (2014).
% https://doi.org/10.1103/PhysRevC.90.061304

\bibitem{Warda12} M. Warda, A. Staszczak and W. Nazarewicz, 
\emph{Fission modes of mercury isotopes}, 
Phys. Rev. C \textbf{86}, 024601 (2012).
% https://doi.org/10.1103/PhysRevC.86.024601

\bibitem{Baran81} A. Baran, K. Pomorski, A. Lukasiak, and A. Sobiczewski, 
\emph{A Dynamic Analysis of Spontaneous-Fission Half-Lives}, 
Nucl. Phys. A \textbf{361}, 83 (1981).

\bibitem{Lemaitre18} J. F. Lemaître, S. Goriely, S. Hilaire and N. Dubray, 
\emph{Microscopic description of the fission path with the Gogny interaction}, 
Phys. Rev. C \textbf{98}, 024623 (2018). 
%https://doi.org/10.1103/PhysRevC.98.024623

\bibitem{Dubray12} N. Dubray and D. Regnier, 
\emph{Numerical search of discontinuities in self-consistent potential energy surfaces}, 
Comp. Phys. Comm. \textbf{183}, 2035-2041 (2012).
% https://doi.org/10.1016/j.cpc.2012.05.001

\bibitem{Bender21} M. Bender and W. Ryssens, in preparation.

\bibitem{Perez09a} S. Perez-Martin and L. M. Robledo, 
\emph{Fission properties of odd-$A$ nuclei in a mean field framework}, 
Int. J. Mod. Phys. E \textbf{18}, 788 (2009).

\bibitem{Heenen16} P.-H. Heenen, B. Bally, M. Bender, and W. Ryssens, 
\emph{Beyond-Mean-Field Correlations and the Description of Superheavy Elements},
Proceedings of the Nobel Symposium on the "Chemistry and Physics of Heavy and Superheavy Elements" (NS160)
held at B{\"a}ckaskog Castle, Kristianstad, Sweden, May 29 - June 3 2016.
D. Rudolph, L.-I. Elding, C. Fahlander and S. {\AA}berg [edts.],
EPJ Web of Conferences \textbf{131} 02001 (2016).

\bibitem{Koh17} M. H. Koh, L. Bonneau, P. Quentin, T. V. N. Hao and H. Wagiran, 
\emph{Fission barriers of two odd-neutron actinide nuclei taking into account the
       Time-reversal symmetry breaking at the mean-field level}, 
Phys. Rev. C \textbf{95}, 014315 (2017).
% https://doi.org/10.1103/PhysRevC.95.014315

\bibitem{Rodriguez17} Rodr\'iguez-Guzm{\'a}n and L. M. Robledo, 
\emph{Microscopic description of fission in odd-mass uranium and plutonium nuclei 
      with the Gogny energy density functional}, 
Eur. Phys J. A \textbf{53}, 245 (2017). 
%https://doi.org/10.1140/epja/i2017-12444-9

\bibitem{Schunck22x} 
 N. Schunck, M. Verriere, G. Potel Aguilar, R. C. Malone, J. A. Silano, A. P. D. Ramirez, and A. P. Tonchev, 
 Microscopic Calculation of Fission Product Yields for Odd-Mass Nuclei, 
 Phys. Rev. C \textbf{107}, 044312 (2023).

\bibitem{Bender00b} M. Bender, K. Rutz, P.-G. Reinhard and J. A. Maruhn, 
\emph{Pairing gaps from nuclear mean-field models}, 
Eur. Phys. J. A \textbf{8}, 59-75 (2000).
%https://doi.org/10.1007/s10050-000-4504-z

\bibitem{Angeli13} I. Angeli and K. P. Marinova, 
\emph{Table of experimental nuclear ground state charge radii: An update},
At. Data Nucl. Data Tables {\bf 99}, 69 (2013).

\bibitem{Samyn04} M. Samyn, S. Goriely, M. Bender and J. M. Pearson, 
\emph{Further explorations of Skyrme-Hartree-Fock-Bogoliubov mass formulas. III. Role of particle-number projection},
Phys. Rev. C \textbf{70}, 044309 (2004). 
%https://doi.org/10.1103/PhysRevC.70.044309

\bibitem{Hunyadi01} M. Hunyadi et al., 
\emph{Excited Superdeformed K$^{\pi}$=0$^+$ Rotational Bands in $\beta$-Vibrational Fission Resonances of $^{240}$Pu}, 
Phys. Lett. B \textbf{505}, 27 (2001).

\bibitem{Singh02} B. Singh, R. Zywina, and R. B. Firestone, 
\emph{Table of Superdeformed Nuclear Bands and Fission Isomers}, 
Nuclear Data Sheets 97, 241 (2002).

\bibitem{Kantele83} J. Kantele, W. St{\"o}ffl, L. E. Ussery, D. J. Decman, E. A. Henry, R. W. Hoff, L. G. Mann, and G. L. Struble, 
\emph{Observation of an $E0$ Isomeric Transition from the $^{238}\mathrm{U}$ Shape Isomer}, 
Phys. Rev. Lett. \textbf{51}, 91 (1983).

\bibitem{Bartel82} J. Bartel, P. Quentin, M. Brack, C. Guet and H.-B. Håkansson,
\emph{Towards a better parametrisation of Skyrme-like effective forces: A critical study of the SkM force},
Nucl. Phys. A \textbf{386}, 79-100 (1982). 
%https://doi.org/10.1016/0375-9474(82)90403-1

\bibitem{Berger89} J. F. Berger, M. Girod and D. Gogny, 
\emph{Constrained Hartree-Fock and beyond},
Nucl. Phys. A \textbf{502}, 85-104 (1989). 
%https://doi.org/10.1016/0375-9474(89)90656-8

\bibitem{Kortelainen12} M. Kortelainen, J. McDonnell, W. Nazarewicz, P.-G. Reinhard, J. Sarich, N. Schunck, M. Stoitsov and S. M. Wild
\emph{Nuclear energy density optimization: Large deformations}, 
Phys. Rev. C \textbf{85}, 024304 (2012). 
%https://doi.org/10.1103/PhysRevC.85.024304

\bibitem{Kortelainen14} M. Kortelainen, J. McDonnell, W. Nazarewicz, E. Olsen, P.-G. Reinhard, J. Sarich, N. Schunck, S. M. Wild, D. Davesne, J. Erler and A. Pastore, 
\emph{Nuclear energy density optimization: Shell structure},
Phys. Rev. C \textbf{89}, 054314 (2014). 
%https://doi.org/10.1103/PhysRevC.89.054314

\bibitem{daCosta22} Ph. Da Costa, M. Bender, K. Bennaceur, J. Meyer, and W. Ryssens,
\emph{in preparation}.

\bibitem{Klupfel09} P. Kl\"upfel, P.-G. Reinhard, T. J. B\"urvenich and J. A. Maruhn, 
\emph{Variations on a theme by Skyrme: A systematic study of adjustments of model parameters},
Phys. Rev. C \textbf{79}, 034310 (2009). 
%https://doi.org/10.1103/PhysRevC.79.034310

\bibitem{Kortelainen10} M. Kortelainen, T. Lesinski, J. Mor\'e, W. Nazarewicz, J. Sarich, N. Schunck, M. V. Stoitsov and S. Wild,
\emph{Nuclear energy density optimization}, 
Phys. Rev. C \textbf{82}, 024313 (2010). 
%https://doi.org/10.1103/PhysRevC.82.024313

\bibitem{Baye86} D. Baye and P.-H. Heenen, 
\emph{Generalised meshes for quantum mechanical problems},
J. Phys. A. Math. Gen. \textbf{19}, 2041 (1986).

\bibitem{Flocard74} H. Flocard, P. Quentin, D. Vautherin, M. Veneroni, and A. K. Kerman, 
\emph{Self-Consistent Calculation of the Fission Barrier of $^{240}$Pu}, 
Nucl. Phys. A \textbf{231}, 176 (1974).

\bibitem{Rutz95} K. Rutz, J. A. Maruhn, P.-G. Reinhard and W. Greiner, 
\emph{Fission barriers and asymmetric ground states in the relativistic mean-field theory}, 
Nucl. Phys. A \textbf{590}, 680-702 (1995).
% https://doi.org/10.1016/0375-9474(95)00192-4

\bibitem{Burvenich04} T. B{\"u}rvenich, M. Bender, J. A. Maruhn and P.-G. Reinhard,
\emph{Systematics of fission barriers in superheavy elements}, 
Phys. Rev. C \textbf{69}, 014307 (2004).
% https://doi.org/10.1103/PhysRevC.69.014307

\bibitem{Bonneau04} L. Bonneau, P. Quentin and D. Sams{\oe}n, 
\emph{Fission barriers of heavy nuclei within a microscopic approach}, 
Eur. Phys. J. A \textbf{21}, 391-406 (2004).
% https://doi.org/10.1140/epja/i2003-10224-x

\bibitem{Samyn05} M. Samyn, S. Goriely and J. M. Pearson, 
\emph{Further explorations of Skyrme-Hartree-Fock-Bogoliubov mass formulas. V. Extension to fission barriers},
Phys. Rev. C \textbf{72}, 044316 (2005).
% https://doi.org/10.1103/PhysRevC.72.044316

\bibitem{Younes09} W. Younes and D. Gogny, 
\emph{Microscopic calculation of $^{240}$Pu scission with a finite-range effective force}, 
Phys. Rev. C \textbf{80}, 054313 (2009).
% https://doi.org/10.1103/PhysRevC.80.054313

\bibitem{Li10} Z. P. Li, T. Nikšić, D. Vretenar, P. Ring and J. Meng, 
\emph{Relativistic energy density functionals: Low-energy collective states of $^{240}$Pu and $^{166}$Er},
Phys. Rev. C \textbf{81}, 064321 (2010).
% https://doi.org/10.1103/PhysRevC.81.064321

\bibitem{Schunck14} N. Schunck, D. Duke, H. Carr and A. Knoll,
\emph{Description of induced nuclear fission with Skyrme energy functionals: Static potential energy surfaces and fission fragment properties}, 
Phys. Rev. C \textbf{90}, 054305 (2014).
% https://doi.org/10.1103/PhysRevC.90.054305

\bibitem{Ryssens19} W. Ryssens, M. Bender and P.-H. Heenen       
\emph{Iterative approaches to the self-consistent nuclear energy density functional problem:
Heavy ball dynamics and potential preconditioning},
Eur. Phys. J. A \textbf{55}, 93 (2019).

\bibitem{Bemis73} C. E. Bemis \emph{et al.}, 
\emph{$E2$ and $E4$ Transition Moments and Equilibrium Deformations in the Actinide Nuclei},
Phys. Rev. C \textbf{8}, 1466 (1973).

\bibitem{Zumbro86} J. D. Zumbro \emph{et al.}, 
\emph{$E2$ and $E4$ Deformations in $^{232}$Th and $^{239,240,242}$Pu}, 
Phys. Lett. B \textbf{167}, 383 (1986).

\bibitem{Ryssens23} W. Ryssens, G. Giacalone, B. Schenke and C. Shen,
Evidence of Hexadecapole Deformation in Uranium-238 at the Relativistic Heavy Ion Collider, 
accepted for publication in Phys. Rev. Lett., arXiv:2302.13617.

\bibitem{RingSchuck} P. Ring and P. Schuck, 
\emph{The Nuclear Many-Body Problem}, 
(Springer-Verlag, New York, 1980).

\bibitem{Bender04} M. Bender, P.-H. Heenen and P. Bonche, 
\emph{Microscopic study of $^{204}$Pu: Mean field and beyond},
Phys. Rev. C \textbf{70}, 054304 (2004). 
%https://doi.org/10.1103/PhysRevC.70.054304

\bibitem{Perez08} S. Perez-Martin and L. M. Robledo, 
\emph{Microscopic justification of the equal filling approximation},
Phys. Rev. C \textbf{78}, 014304 (2008).
% https://doi.org/10.1103/PhysRevC.78.014304

\bibitem{Schunck10} 
N. Schunck, J. Dobaczewski, J. McDonnell, J. Moré, W. Nazarewicz, J. Sarich, and M. V. Stoitsov, 
\emph{One-Quasiparticle States in the Nuclear Energy Density Functional Theory}, 
Phys. Rev. C 81, 024316 (2010).

\bibitem{Kowal10} M. Kowal, P. Jachimowicz and A. Sobiczewski, 
\emph{Fission barriers for even-even superheavy nuclei}, 
Phys. Rev. C \textbf{82}, 014303 (2010).
% https://doi.org/10.1103/PhysRevC.82.014303

\bibitem{Jachim17} P. Jachimowicz, M. Kowal and J. Skalski,
\emph{Adiabatic fission barriers in superheavy nuclei},
Phys. Rev. C \textbf{95}, 014303 (2017). 
%https://doi.org/10.1103/PhysRevC.95.014303

\bibitem{Delaroche06} J. P. Delaroche, M. Girod, H. Goutte and J. Libert,
\emph{Structure properties of even-even actinides at normal and super deformed shapes analysed using the Gogny force}, 
Nucl. Phys. A \textbf{771}, 103-168 (2006).
% https://doi.org/10.1016/j.nuclphysa.2006.03.004

\bibitem{Lu2012} B.-N. Lu, E.-G. Zhao, and S.-G. Zhou, 
\emph{Potential Energy Surfaces of Actinide Nuclei from a Multidimensional Constrained Covariant Density Functional Theory: Barrier Heights and Saddle Point Shapes}, 
Phys. Rev. C \textbf{85}, 011301 (2012).

\bibitem{Pashkevich69} V. V. Pashkevich, 
\emph{The Energy of Non-Axial Deformation of Heavy Nuclei},
Nucl. Phys. A \textbf{133}, 400 (1969).

\bibitem{Larsson72} S. E. Larsson, I. Ragnarsson, and S. G. Nilsson, 
\emph{Fission Barriers and the Inclusion of Axial Asymmetry}, 
Phys. Lett. B \textbf{38}, 269 (1972).

\bibitem{Gotz72} U. G{\"o}tz, H. C. Pauli, and K. Junker, 
\emph{Influence of axially asymmetric distortions on fission barriers}, 
Phys. Lett. B \textbf{39}, 436 (1972).

\bibitem{Dutta00} A. K. Dutta, J. M. Pearson and F. Tondeur, 
\emph{Triaxial nuclei calculated with the extended Thomas-Fermi plus Strutinsky integral (ETFSI) method}, 
Phys. Rev. C \textbf{61}, 054303 (2000). 
%https://doi.org/10.1103/PhysRevC.61.054303

\bibitem{Moller09} P. Möller, A. J. Sierk, T. Ichikawa, A. Iwamoto, R. Bengtsson, H. Uhrenholt, and S. Åberg, 
\emph{Heavy-Element Fission Barriers}, 
Phys. Rev. C \textbf{79}, 064304 (2009).

\bibitem{Jachim12} P. Jachimowicz, M. Kowal and J. Skalski,
\emph{Secondary fission barriers in even-even actinide nuclei},
Phys. Rev. C \textbf{85}, 034305 (2012). 
%https://doi.org/10.1103/PhysRevC.85.034305

\bibitem{Sadhukhan13} J. Sadhukhan, K. Mazurek, A. Baran, J. Dobaczewski, W. Nazarewicz, and J. A. Sheikh, 
\emph{Spontaneous Fission Lifetimes from the Minimization of Self-Consistent Collective Action}, 
Phys. Rev. C \textbf{88}, 064314 (2013).

\bibitem{Giuliani18} S. A. Giuliani, G. Martínez-Pinedo and L. M. Robledo, 
\emph{Fission properties of superheavy nuclei for r-process calculations},
Phys. Rev. C \textbf{97}, 034323 (2018). 
%https://doi.org/10.1103/PhysRevC.97.034323

\bibitem{Matsu10} K. Matsuyanagi, M. Matsuo, T. Nakatsukasa, N. Hinohara, and K. Sato, 
\emph{Open Problems in the Microscopic Theory of Large-Amplitude Collective Motion}, 
J. Phys. G: Nucl. Part. Phys. \textbf{37}, 064018 (2010).

\bibitem{Baran11} A. Baran, J. A. Sheikh, J. Dobaczewski, W. Nazarewicz and A. Staszczak, 
\emph{Quadrupole collective inertia in nuclear fission: Cranking approximation}, 
Phys. Rev. C \textbf{84}, 054321 (2011). 
%https://doi.org/10.1103/PhysRevC.84.054321

\bibitem{Washiyama21} K. Washiyama, N. Hinohara, and T. Nakatsukasa, 
\emph{Finite-amplitude method for collective inertia in spontaneous fission}, 
Phys. Rev. C \textbf{103}, 014306 (2021).

\bibitem{Jachim20} P. Jachimowicz, M. Kowal, and J. Skalski,
\emph{Static Fission Properties of Actinide Nuclei}, 
Phys. Rev. C \textbf{101}, 014311 (2020).

\bibitem{Erler12} J. Erler, K. Langanke, H. P. Loens, G. Martínez-Pinedo, and P.-G. Reinhard, 
\emph{Fission Properties for r-process Nuclei}, 
Phys. Rev. C \textbf{85}, 025802 (2012).

\bibitem{Taninah20} A. Taninah, S. E. Agbemava, and A. V. Afanasjev, 
\emph{Covariant Density Functional Theory Input for r -Process Simulations in Actinides and Superheavy Nuclei: The Ground State and Fission Properties},
Phys. Rev. C \textbf{102}, 054330 (2020).

%%%%%%%%%%%%%%%%%%%%%%%%%%%%%%%%%%%%%%%%%%%%%%%%%%%%%%%%%%%%%%%%%%%%%%%%%%%%%%%%
% Triaxially deformed fission barriers

\bibitem{Jachimowicz21} P. Jachimowicz, M. Kowal, and J. Skalski, 
\emph{Properties of Heaviest Nuclei with 98$\leq$Z$\leq$126 and 134 $\leq$ N $\leq$ 192}, 
Atomic Data and Nuclear Data Tables \textbf{138}, 101393 (2021).

\bibitem{Lalazissis97} G. A. Lalazissis, J. König, and P. Ring, 
\emph{New Parametrization for the Lagrangian Density of Relativistic Mean Field Theory}, 
Phys. Rev. C \textbf{55}, 540 (1997).

\bibitem{Lalazissis09} G. A. Lalazissis, S. Karatzikos, R. Fossion, D. P. Arteaga, A. V. Afanasjev, and P. Ring, 
\emph{The Effective Force NL3 Revisited}, 
Phys. Lett. B \textbf{671}, 36 (2009).

\bibitem{Muntian01} I. Muntian, Z. Patyk, and A. Sobiczewski, 
\emph{Sensitivity of Calculated Properties of Superheavy Nuclei to Various Changes}, 
Acta Physica Polonica B \textbf{32}, 691 (2001).

\bibitem{Niksic08} T. Nikšić, D. Vretenar, and P. Ring, 
\emph{Relativistic Nuclear Energy Density Functionals: Adjusting Parameters to Binding Energies}, 
Phys. Rev. C \textbf{78}, 034318 (2008).

\bibitem{Karatzikos10} S. Karatzikos, A. V. Afanasjev, G. A. Lalazissis, and P. Ring, 
\emph{The Fission Barriers in Actinides and Superheavy Nuclei in Covariant Density Functional Theory}, 
Phys. Lett. B \textbf{689}, 72 (2010).

\bibitem{Zhao10} P. W. Zhao, Z. P. Li, J. M. Yao, and J. Meng, 
\emph{New Parametrization for the Nuclear Covariant Energy Density Functional with a Point-Coupling Interaction}, 
Phys. Rev. C \textbf{82}, 054319 (2010).

\bibitem{Goriely08b} S. Goriely, S. Hilaire and A. J. Koning, 
\emph{Improved Microscopic Nuclear Level Densities within the Hartree-Fock-Bogoliubov plus Combinatorial Method}, 
Phys. Rev. C \textbf{78}, 064307 (2008).

\bibitem{Schmidt16} K.-H. Schmidt, B. Jurado, C. Amouroux, and C. Schmitt, 
\emph{General Description of Fission Observables: GEF Model Code}, 
Nucl. Data Sheets \textbf{131}, 107 (2016).

\bibitem{Lemaitre19} J.-F. Lemaître, S. Goriely, S. Hilaire, and J.-L. Sida, 
\emph{Fully Microscopic Scission-Point Model to Predict Fission Fragment Observables},
Phys. Rev. C \textbf{99}, 034612 (2019).

\bibitem{Lemaitre21} J.-F. Lemaître, S. Goriely, A. Bauswein, and H.-T. Janka, 
\emph{Fission Fragment Distributions and Their Impact on the r-Process Nucleosynthesis in Neutron Star Mergers}, 
Phys. Rev. C \textbf{103}, 025806 (2021).

\bibitem{Hao12} T. V. N. Hao, P. Quentin and L. Bonneau,
\emph{Parity restoration in the highly truncated diagonalization approach: 
      Application to the outer fission barrier of $^{240}$Pu}, 
Phys. Rev. C \textbf{86}, 064307 (2012). 
%https://doi.org/10.1103/PhysRevC.86.064307

\bibitem{Bernard19} R. Bernard, S. A. Giuliani, and L. M. Robledo, 
\emph{Role of Dynamic Pairing Correlations in Fission Dynamics},
Phys. Rev. C \textbf{99}, 064301 (2019).

\bibitem{Marevic20} P. Marević and N. Schunck, 
\emph{Fission of $^{240}$Pu with Symmetry-Restored Density Functional Theory}, 
Phys. Rev. Lett. \textbf{125}, 102504 (2020).

\bibitem{Dobaczewski19} J. Dobaczewski, 
\emph{Density Functional Theory for Nuclear Fission -- a Proposal}, 
arXiv:1910.03924.

\end{thebibliography}
\end{document}